\documentstyle[mncite]{mn} 

\def\dosingle#1::::{#1}  \def\dodouble#1::::{ } 

\def\nice#1::::{#1}    \def\subm#1::::{}   
\newcommand\zzz[2]{#2}  

\newcommand\yyy[2]{#2}  

\def\datadir{.}

\def\k{\mbox{\rm\,km\,s$^{-1}$}}

\def\kpc{\mbox{\rm kpc}}
\def\hMpc{\mbox{$h^{-1}\mbox{\rm Mpc}$}}
\def\centreline{\centerline}
\def\.{{\cdot}}
\def\gtapprox{\,\lower.6ex\hbox{$\buildrel >\over \sim$} \, }
\def\ltapprox{\,\lower.6ex\hbox{$\buildrel <\over \sim$} \, }
\def\sun{\odot}
\def\msls{\,M_\odot L_\odot^{-1}}
\def\e{ {\scriptstyle \times} 10^}
\def\etal{\mbox{et\,al.}}
\def\arcs{\ifmmode {'' }\else $'' $\fi}     
\def\arcm{\ifmmode {' }\else $' $\fi}     
\def\deg{\ifmmode^\circ\else$^\circ$\fi}    

\def\MM{{\cal M}}

\def\frtoday{le\space\number\day\space\ifcase\month\or
  janvier\or f\'evrier\or mars\or avril\or mai\or juin\or
  juillet\or ao\^ut\or septembre\or octobre\or novembre\or d\'ecembre\fi\space \number\year}

\def\rthresh{r_{\protect\mbox{\rm \small thresh}}}
\def\nlevels{n_{\mbox{\rm \small levels}}}
\def\nlev_tiny{n_{\mbox{\rm \tiny levels}}} 
\def\rinc{r_{\mbox{\rm \small inc}}}
\def\tdyn{t_{\mbox{\rm \small dyn}}}
\def\tburst{t^{\mbox{\rm \small burst}}}

\def\Mtot{M^{\mbox{\rm \small halo}}}
\def\Mgas{M^{\mbox{\rm \small gas}}}
\def\Mgal{M^{\mbox{\rm \small gal}}}

\def\MMtot{\MM_{\mbox{\rm \small tot}}}

\def\MMlum{\MM_{\mbox{\rm \small lum}}}

\def\disk{_{\mbox{\rm \small disk}}}
\def\halo{_{\mbox{\rm \small halo}}}
\def\IIIaJ{_{\mbox{\rm \small IIIaJ}}}

\def\imax{i^{\mbox{\rm \small max}}}
\def\imaxtiny{i^{\mbox{\rm \tiny max}}}

\def\dvir{\delta_{\mbox{\rm \small vir}}}

\def\rthstrut{ \mbox{\rule[-1.5ex]{0cm}{1ex}} }

\def\jref#1;;#2;;#3;;#4 {#1, {#2, }{#3, }#4}

\def\bref#1;;#2;;#3;;#4;;#5 {#1, {in #2 }(#3: #4)\  #5}
\def\psubref#1;;#2;;#3;; {#1. {\em#2, }submitted to {\em#3}}

\def\pserref#1;;#2;;#3 {#1. {\em#2, }{\bf#3}}

\def\unpub#1;;#2;; {#1. {\em#2}}

\def\pub #1;;#2;;#3;;#4;;#5 {#1. {\em#2, }{\ \em#3,  }
{\bf#4, }#5}

\def\prep #1;;#2;;{#1. {\em#2} }

\newcommand\joref[5]{#1, #5, {#2, }{#3, } #4}

\newcommand\epref[3]{#1, #3, #2}

\def\MNRAS{M.N.R.A.S.}
\def\apj{Ap.J.}                 
\def\apjs{Ap.J.Supp.}                 
\def\aj{A.J.}                       
\def\aanda{A.\&A.}            
\def\pasj{P.A.S.J.}

\dodouble \documentstyle[doublespacing,mncite]{mn} ::::

\nice \input epsf.tex ::::

\begin{document}

\nice \newcommand\pixfig[1]{#1} ::::  
\nice \newcommand\bigfig[1]{}   :::: 

\title[Merging History Trees]{Merging History Trees of Dark Matter Haloes:            a Tool for Exploring Galaxy Formation Models}

\def\IAP{Institut d'Astrophysique de Paris, 98bis Bd Arago, F-75.014 Paris,
France}
\def\sussex{Astronomy Centre, 
University of Sussex, Falmer, Brighton,
BN1~9QH, United Kingdom}
\def\mssso{Mt Stromlo \& Siding Spring Observatories, 
Locked Bag, Weston Creek P.O., ACT 2611, Australia}
\def\naoj{National Astronomical Observatory, Mitaka, Tokyo 181, Japan}
\def\eso{European Southern Observatory, D-85748, 
Garching bei M\"unchen, Germany}

\author[B.F.Roukema \etal]{B.F.Roukema$^{1,2,3,4}$, B.A.Peterson$^1$, 
P.J.Quinn$^{1,5}$ \& B.Rocca-Volmerange$^2$\\
{$^1$\mssso}\\{$^2$\IAP}\\{$^3$\sussex}\\{$^4$\naoj}\\{$^5$\eso}\\
{Email: roukema@iap.fr, peterson@mso.anu.edu.au, pjq@eso.org, rocca@iap.fr} }

\def\today{\frtoday}
\maketitle

\dodouble \clearpage ::::

\begin{abstract}
A method of deriving and using merging history trees 
of dark matter galaxy haloes 
directly from pure gravity N-body simulations is presented. 
This combines the full non-linearity of N-body simulations with the
flexibility of the semi-analytical approach.

Merging history trees derived from 
 power-law initial perturbation spectrum simulations 
(for indices $n=-2$ and $n=0$) by \cite{Warr92}~(1992) are shown. 
As an example of a galaxy formation model, these are combined with 
evolutionary stellar population synthesis, via 
simple scaling laws for star formation rates, 
showing that if most star formation occurs during merger-induced bursts, 
then a nearly flat 
faint-end slope of the galaxy luminosity function 
may be obtained in certain cases.

Interesting properties of hierarchical halo formation are noted: 
(1) In a given model, merger rates may vary widely between individual 
haloes, and typically 20\%$\sim$30\% of a halo's mass may be due 
to infall of uncollapsed material. 
(2) Small mass haloes continue to form at recent times: as expected, 
the existence of 
young, low redshift, 
low metallicity galaxies (e.g., \cite{Lipo97}~1997) is consistent with 
hierarchical galaxy formation models. 
(3) For $n=-2,$ the halo spatial correlation function can have a very high 
initial bias due to the high power on large scales. 

\end{abstract}


\def\fprof{
\begin{figure}
\centering 
\nice \centreline{\epsfxsize=7cm
\zzz{ \epsfbox[25 18 578 574]{"`gunzip -c \datadir/profile.ps.gz"}}
  {\epsfbox[25 18 578 574]{"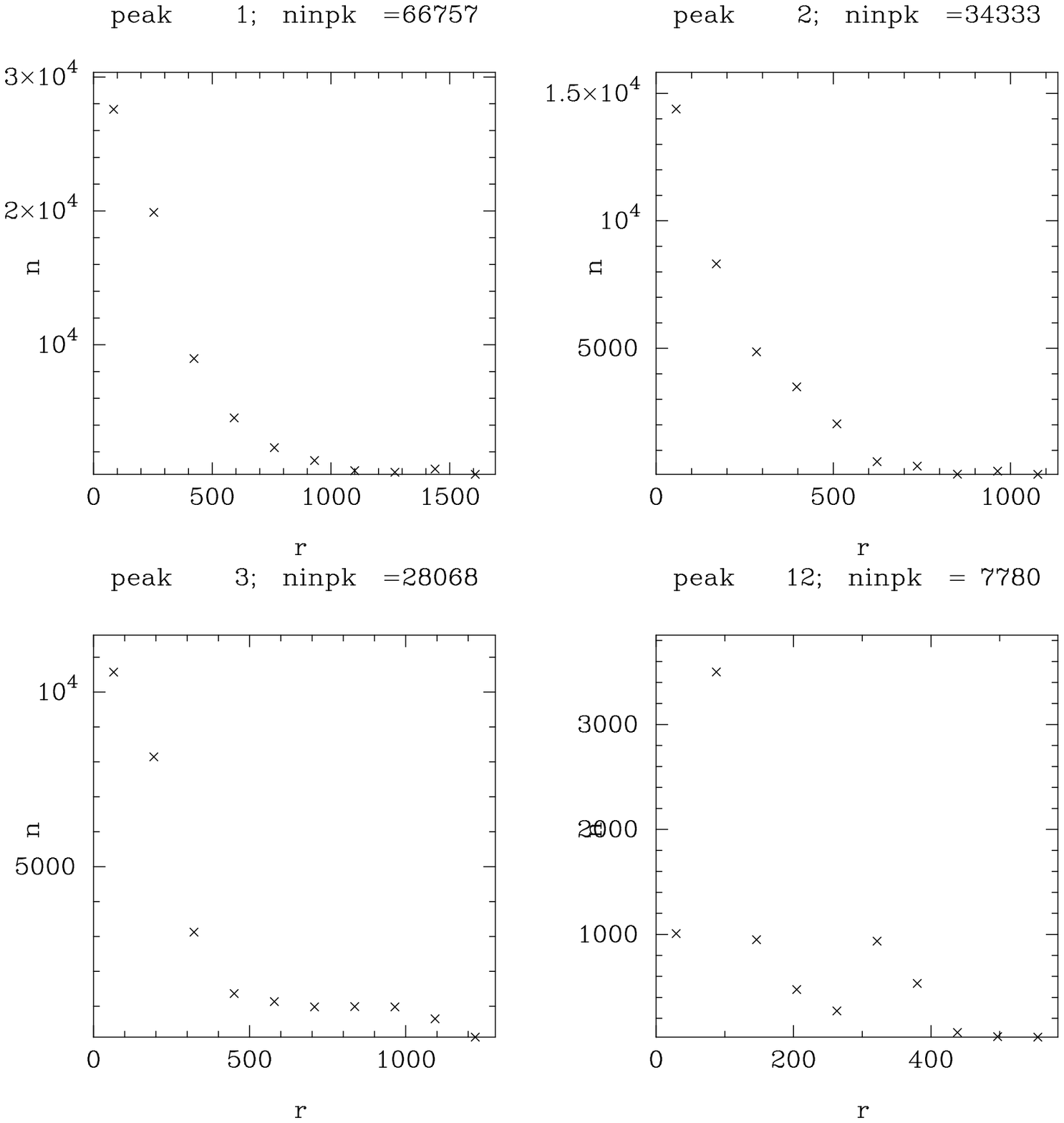"}}
}
::::
\subm \vspace{7cm} ::::
\caption[Halo profiles (a)]{ 
\label{f-profile} Halo profiles: numbers of particles in spherical shells 
	for some of the most massive haloes
	(``peak'' number = $1-3,12$ as labelled) 
	of time stage 615 in the $n=-2$ model for  $\rthresh =5.$ 
	The total number of particles in each peak is labelled ``ninpk''.}
\end{figure}
}

\def\mhplots{
\begin{figure}
\centering 
\nice \centreline{\epsfxsize=7cm
\bigfig{
\yyy{ \epsfbox[25 18 578 574]{"`gunzip -c \datadir/mhn-2b.r5.1_5.ps.gz"}} 
{ \epsfbox[25 18 578 574]{"mhn-2b.r5.1_5.ps"}} 
}
\pixfig{
\yyy{ \epsfbox[25 18 578 574]{"`gunzip -c \datadir/mhn-2b.r5.1_5.2.ps.gz"}} 
{ \epsfbox[25 18 578 574]{"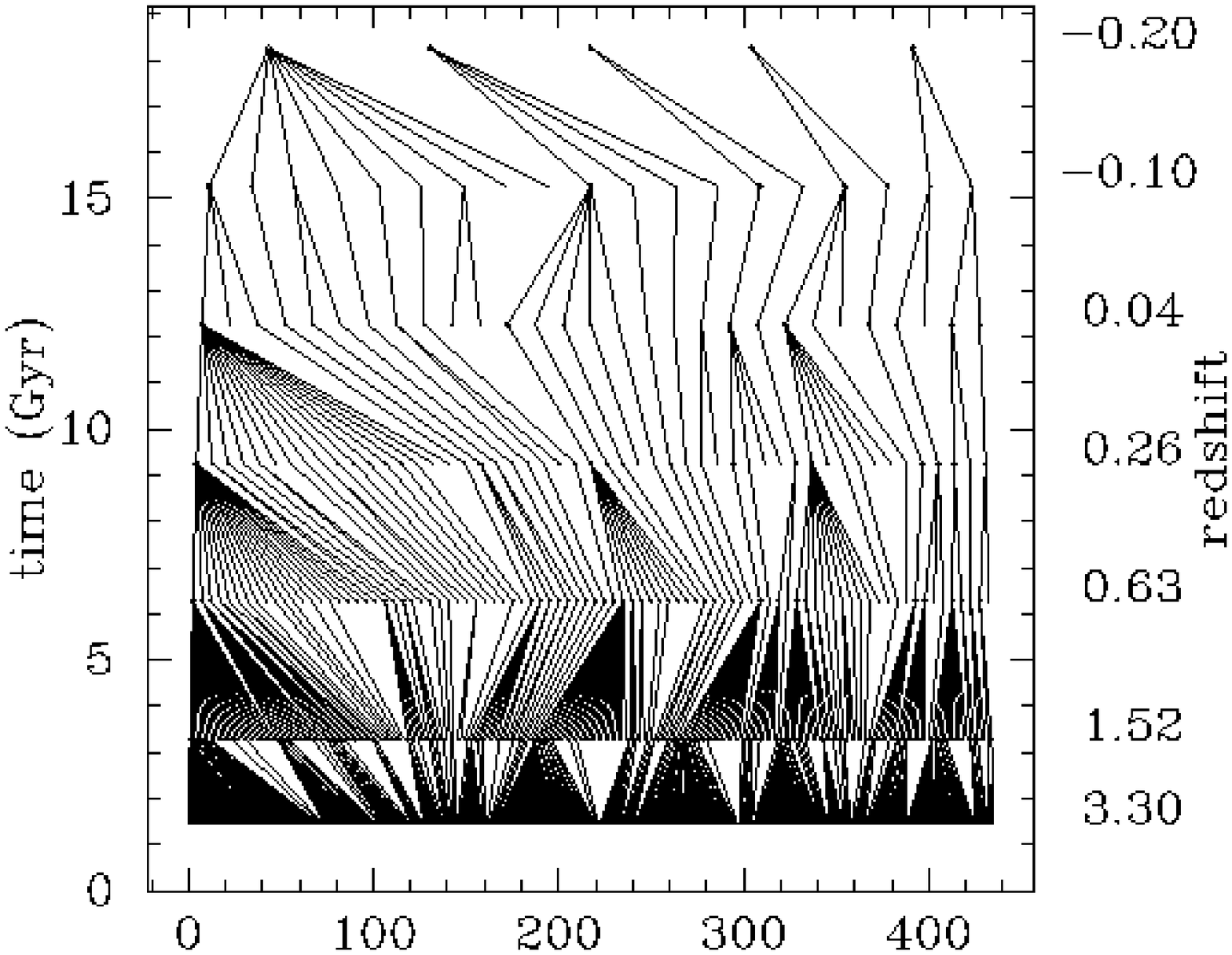"}} 
}}
::::
\subm \vspace{7cm} ::::
\caption[Merging History: $n=-2,\, \protect\rthresh =5$, haloes $1 - 5.$
]{ 
	\label{f-mhn-2b.r5.1_5}   
Merging History: $n=-2,\, \protect\rthresh =5$, 
haloes $1 - 5.$  This and the following plots show haloes detected at
different points in space-time connected according to the criterion 
described in \S\protect\ref{s-tree}, 
i.e., showing which haloes merge into which.
The horizontal axis separates individual haloes,
while the vertical axis indicates time/redshift. 
(Negative redshifts indicate future times.)
Circles indicate 
haloes, with radii a logarithmic transformation of the haloes masses 
(for display purposes, the specific transformation differs
between separate plots) and line segments indicating
the merging connections. The haloes at the latest time stage, and the set
of predecessors of any halo, are ordered by mass decreasing to the right.
Numbering on the horizontal axis indicates the maximum number of haloes
in the figure for any time stage.}
\end{figure}

\begin{figure}
\centering 
\nice \centreline{\epsfxsize=7cm
\bigfig{
\yyy{ 
\epsfbox[25 18 578 574]{"`gunzip -c \datadir/mhn-2b.r1000.1_5.ps.gz"}} 
{ \epsfbox[25 18 578 574]{"mhn-2b.r1000.1_5.ps"}} 
}
\pixfig{
\yyy{ 
\epsfbox[25 18 578 574]{"`gunzip -c \datadir/mhn-2b.r1000.1_5.2.ps.gz"}} 
{ \epsfbox[25 18 578 574]{"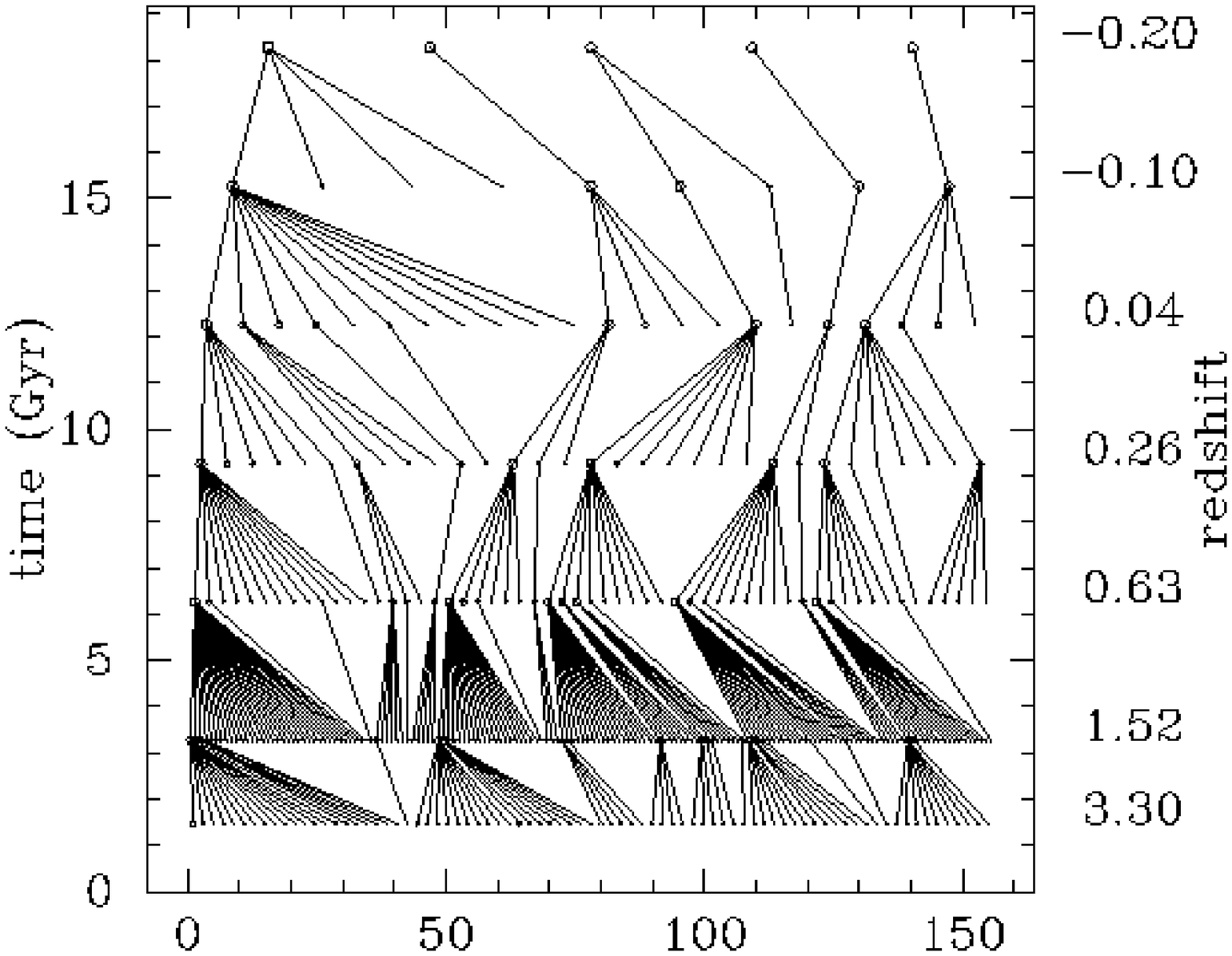"}} 
}}
::::
\subm \vspace{7cm} ::::
\caption{ \label{f-mhn-2b.r1000.1_5} 
Merging History: $n=-2,\, \protect\rthresh =1000$, 
haloes $1-5$ }
\end{figure}

\begin{figure}
\centering 
\nice \centreline{\epsfxsize=7cm
\bigfig{
\yyy{ 
\epsfbox[25 18 578 574]{"`gunzip -c \datadir/mhn-2b.r5.50_60.ps.gz"}} 
{ \epsfbox[25 18 578 574]{"mhn-2b.r5.50_60.ps"}} 
}
\pixfig{
\yyy{ 
\epsfbox[25 18 578 574]{"`gunzip -c \datadir/mhn-2b.r5.50_60.2.ps.gz"}} 
{ \epsfbox[25 18 578 574]{"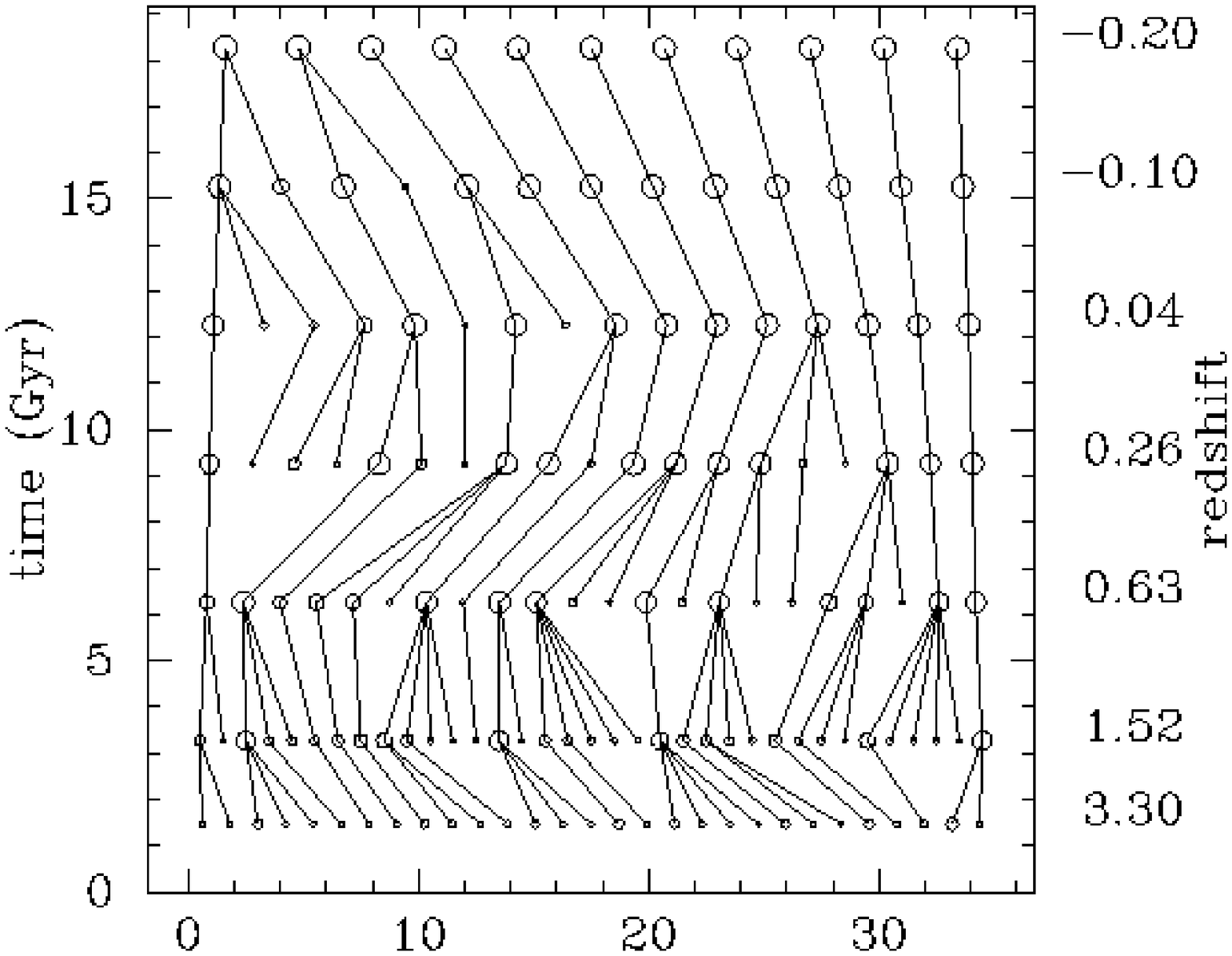"}} 
}}
::::
\subm \vspace{7cm} ::::
\caption{ \label{f-mhn-2b.r5.50_60}
Merging History: $n=-2,\, \protect\rthresh =5$, haloes $50-60$ }
\end{figure}

\begin{figure}
\centering 
\nice \centreline{\epsfxsize=7cm
\zzz{ 
\epsfbox[25 18 578 574]{"`gunzip -c \datadir/mhn-2b.r5.400_410.ps.gz"}} 
{ \epsfbox[25 18 578 574]{"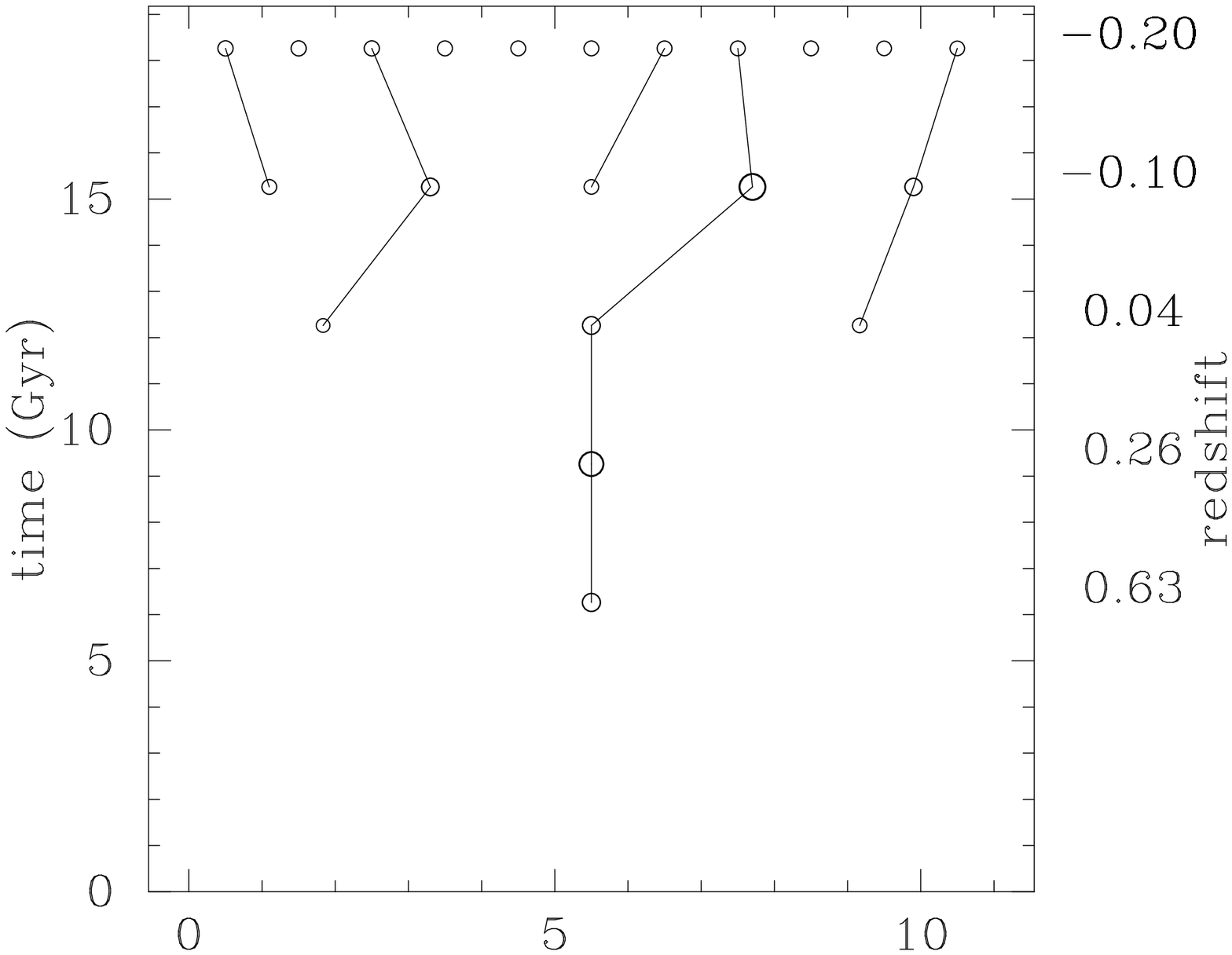"}} 
}
::::
\subm \vspace{7cm} ::::
\caption{ \label{f-mhn-2b.r5.400_410}
Merging History: $n=-2,\, \protect\rthresh =5$, 
haloes $400-410$ }
\end{figure}

} 

\def\fmratios{
\begin{figure}
\centering 
\nice \centreline{\epsfxsize=11cm
\zzz{ 
\epsfbox[25 18 528 774]{"`gunzip -c \datadir/rntn1.n-2b.r5.ps.gz"}} 
{ \epsfbox[25 18 528 774]{"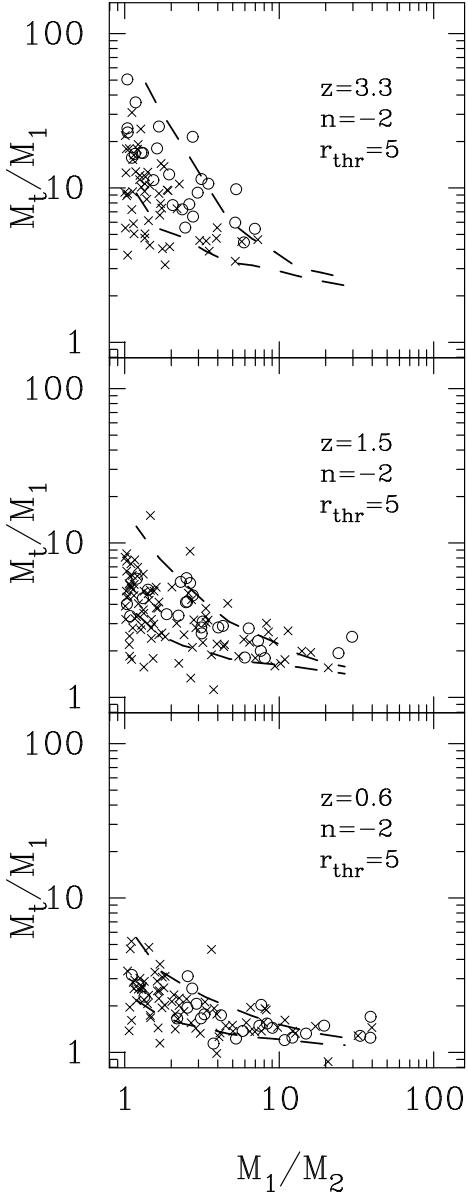"}} 
}
::::
\subm \vspace{15cm} ::::
\caption[Mass ratios $N_t/N_1,$ $N_1/N_2$.]{ \label{f-rntn1_r5}
Mass ratios of halo at final time stage ($M_t$) and the masses of 
the first and second most massive progenitor haloes ($M_1,$ $M_2$ resp.) 
of that halo at a redshift $z$, for $n=-2$ and $\protect\rthresh=5$. 
Circles (crosses) 
are for haloes more (less) massive than $10^{11}M_{\sun}.$ The dashed lines
outline the region in the diagram covered by the 
semi-analytical haloes of \protect\cite{KW93}~(1993) for a CDM initial 
fluctuation spectrum and a bias factor of $b=2\.5$ [$\log_{10}(M_t/M_1)$ 
values are linearly interpolated/extrapolated in $z$ from those in Fig.~4 of 
\protect\cite{KW93}].
}
\end{figure}

\begin{figure}
\centering 
\nice \centreline{\epsfxsize=11cm
\zzz{ 
\epsfbox[25 18 528 774]{"`gunzip -c \datadir/rntn1.n-2b.r1000.ps.gz"}} 
{ \epsfbox[25 18 528 774]{"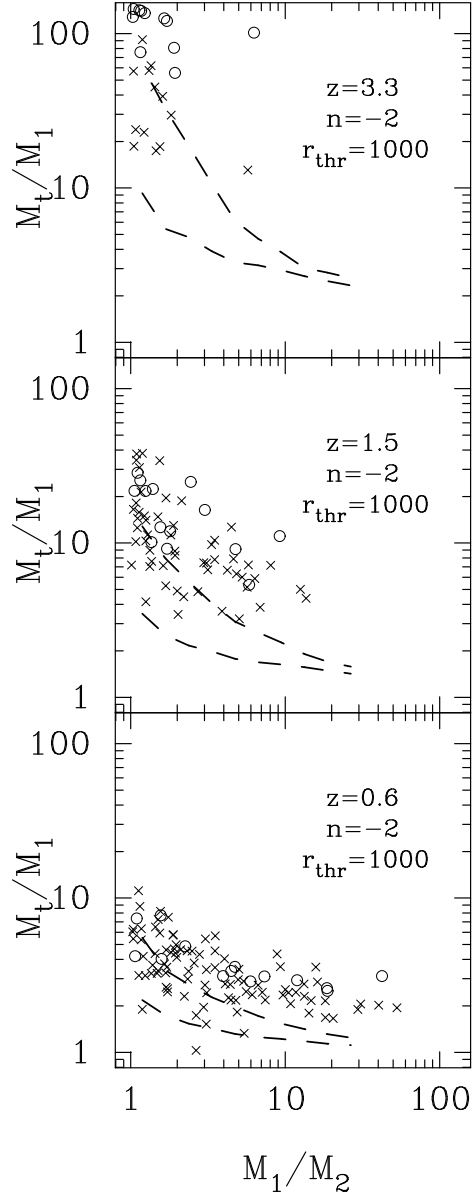"}} 
}
::::
\subm \vspace{15cm} ::::
\caption{ \label{f-rntn1_r1000}
Progenitor mass ratios as per Fig.~\protect\ref{f-rntn1_r5}, for 
$\protect\rthresh=1000.$
}
\end{figure}
}  

\def\fmfns{
\begin{figure}
\centering 
\nice \centreline{\epsfxsize=7cm
 \zzz{ \epsfbox[25 18 578 574]{"`gunzip -c \datadir/mfn-2b.r5.ps.gz"}}
{ \epsfbox[25 18 578 574]{"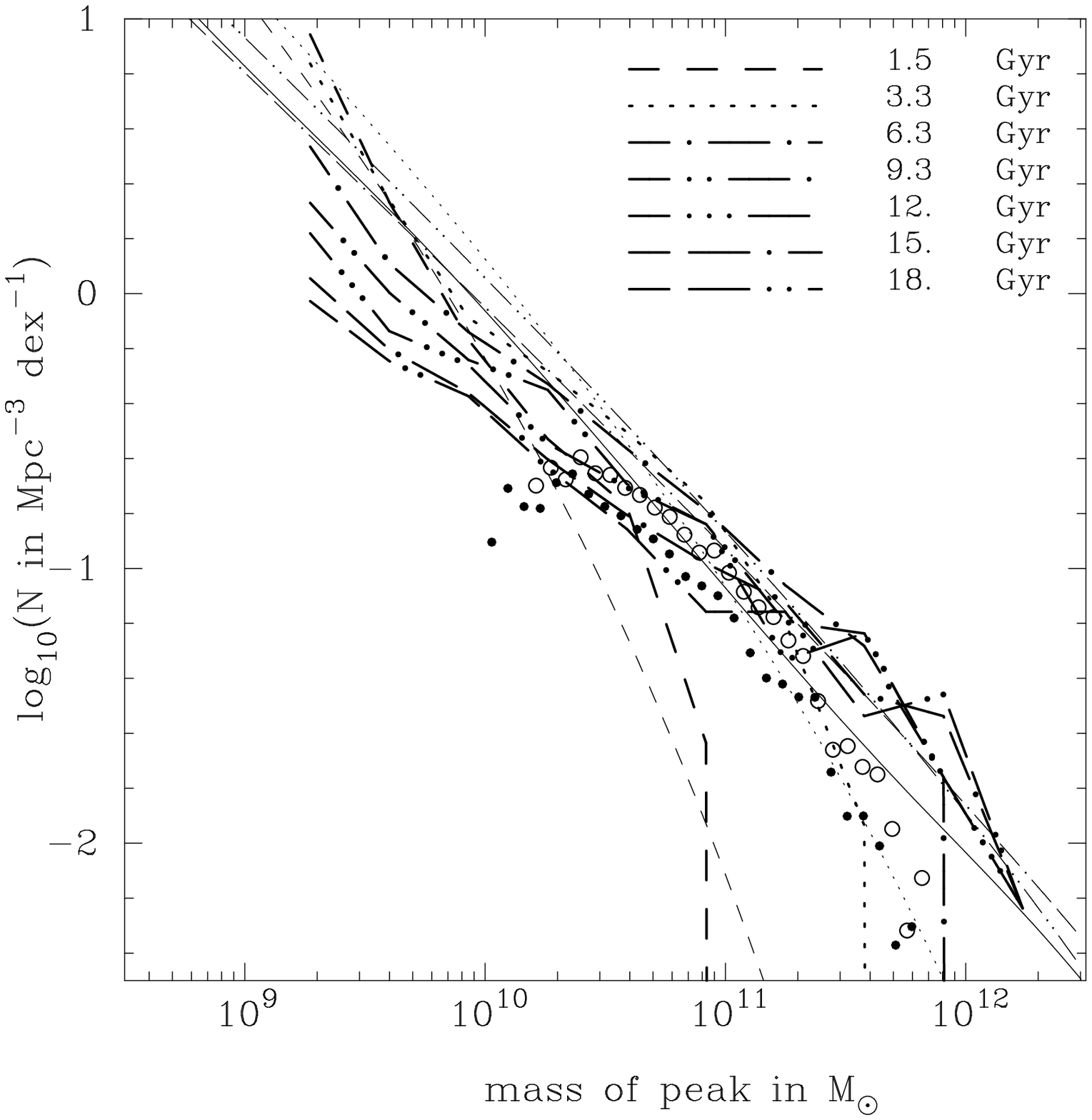"}}
}
::::
\subm \vspace{7cm} ::::
\caption[Mass functions for $n=-2 $ model for $\protect\rthresh =5.$]{ 
 \label{f-mfn-2br5} Mass functions for the $n=-2 $ model 
for $\protect\rthresh =5.$ 
Thick lines are for this work at the time stages labelled. 
For comparison with other results  
(\S\protect\ref{s-cfother}), 
the Press \& Schechter formula (\cite{LC93}~1993; thin lines; 
$\delta_{c0} = \protect\dvir \equiv 1\.686$) and  
the models of \protect\cite{LS91}~(1991) (thin solid line)
and of \protect\cite{BoMy96}~(1996) (hollow and solid circles for 
spherical and ellipsoidal internal dynamics resp.) 
for a CDM initial fluctuation spectrum are shown.
}
\end{figure}

\begin{figure}
\centering 
\nice \centreline{\epsfxsize=7cm
\zzz{   \epsfbox[25 18 578 574]{"`gunzip -c \datadir/mfn0b.r5.ps.gz"}}
 {  \epsfbox[25 18 578 574]{"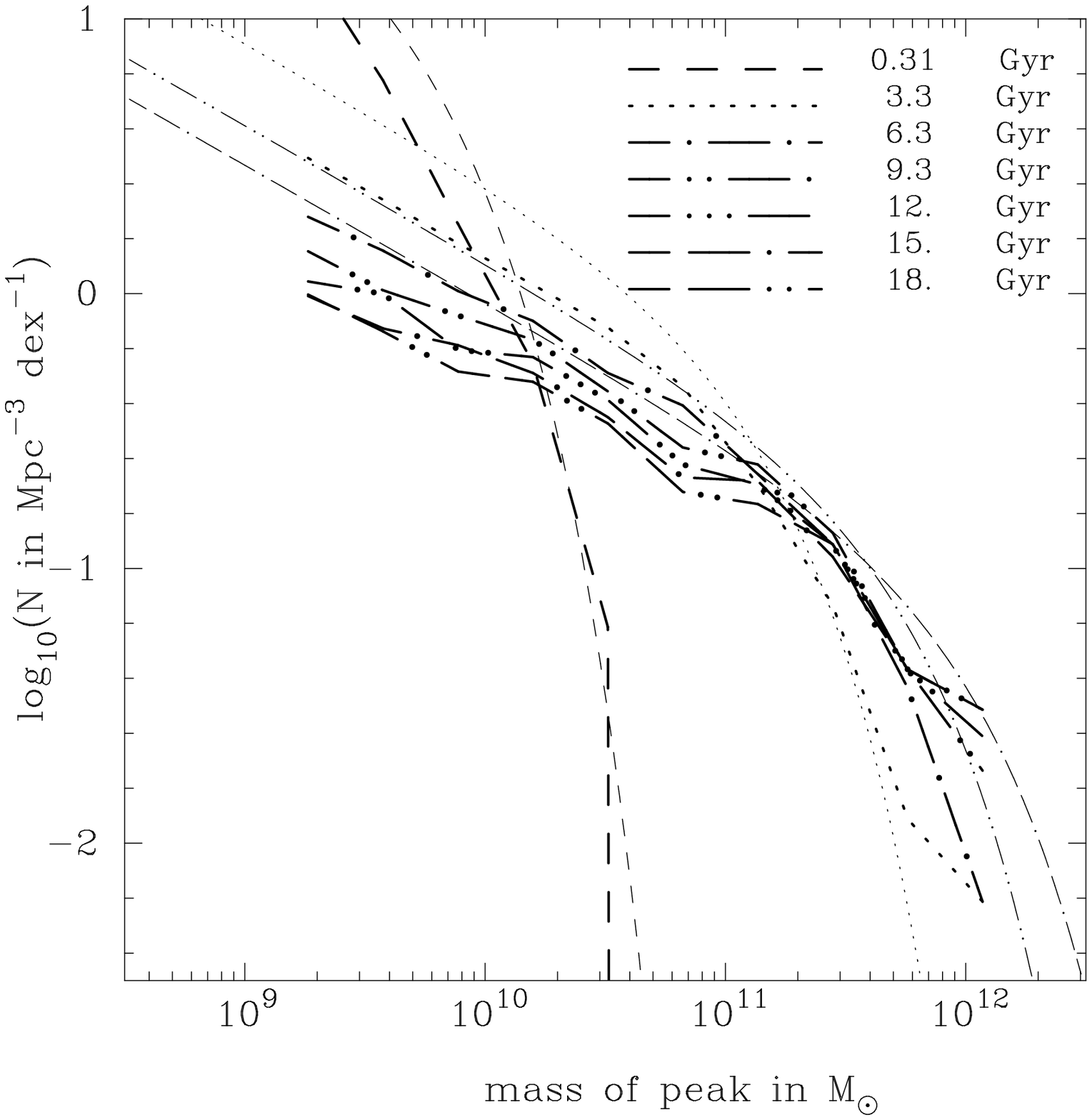"}}
}
::::
\subm \vspace{7cm} ::::
\caption[Mass functions for $n=0 $ model for $\protect\rthresh =5.$]{ 
\label{f-mfn0br5} Mass functions for $n=0 $ model for $\protect\rthresh =5.$
Curve styles are as for Fig.~\ref{f-mfn-2br5}. 
}
\end{figure}

\begin{figure}
\centering 
\nice \centreline{\epsfxsize=7cm
 \zzz{ \epsfbox[25 18 578 574]{"`gunzip -c \datadir/mfn-2b.r1000.ps.gz"}}
{ \epsfbox[25 18 578 574]{"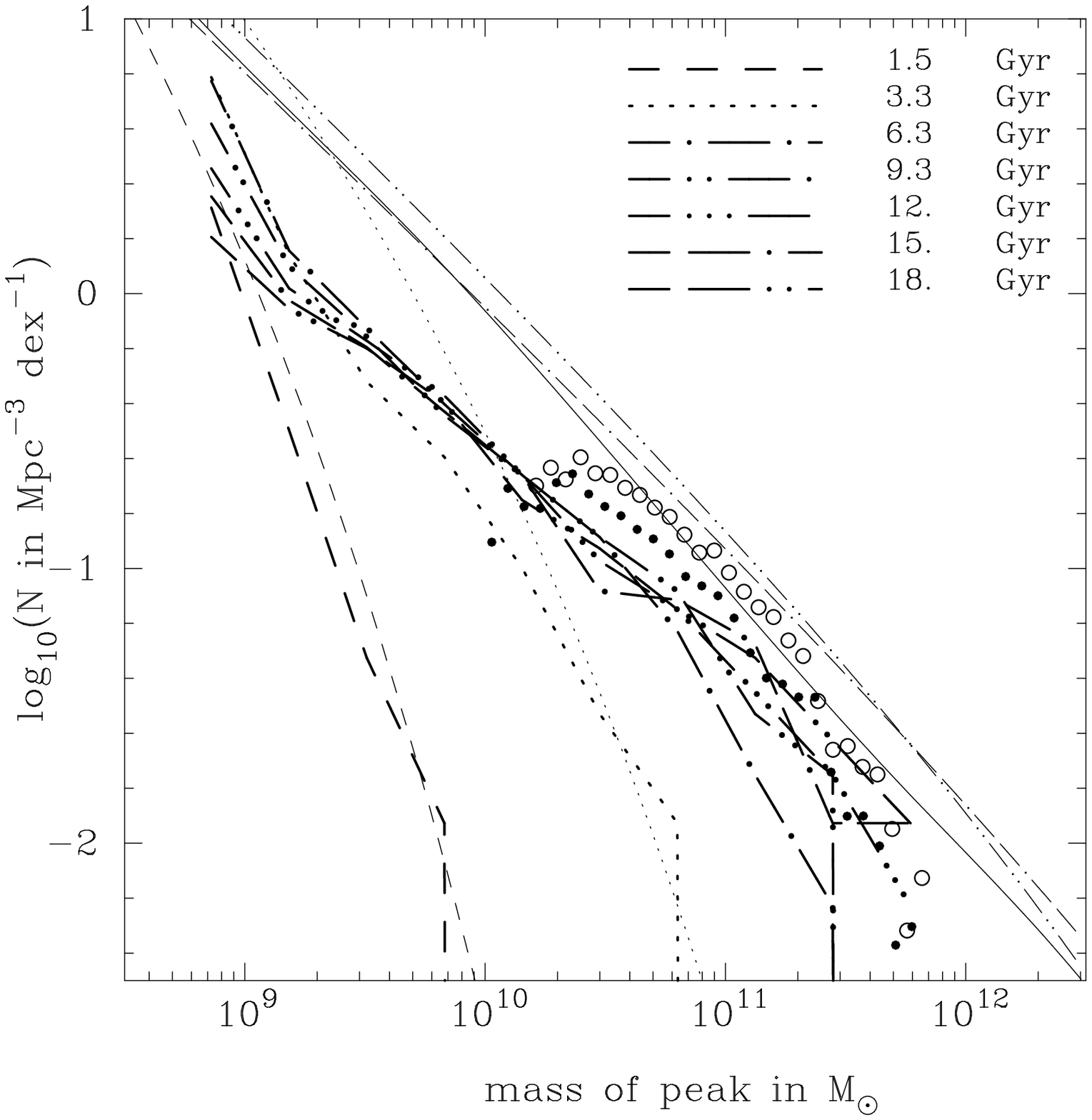"}}
}
::::
\subm \vspace{7cm} ::::
\caption[Mass functions for $n=-2 $ model for $\protect\rthresh =1000.$]{ 
 \label{f-mfn-2br1000} Mass functions for $n=-2 $ model 
for $\protect\rthresh =1000.$ Curves are as for Fig.~\ref{f-mfn-2br5} 
except that $\delta_{c0} = 2 \protect\dvir$ in the PS formula for the 
$1\.5$~Gyr and $3\.3$~Gyr time steps.} 
\end{figure}

\begin{figure}
\centering 
\nice \centreline{\epsfxsize=7cm
\zzz{   \epsfbox[25 18 578 574]{"`gunzip -c \datadir/mfn0b.r1000.ps.gz"}}
 {  \epsfbox[25 18 578 574]{"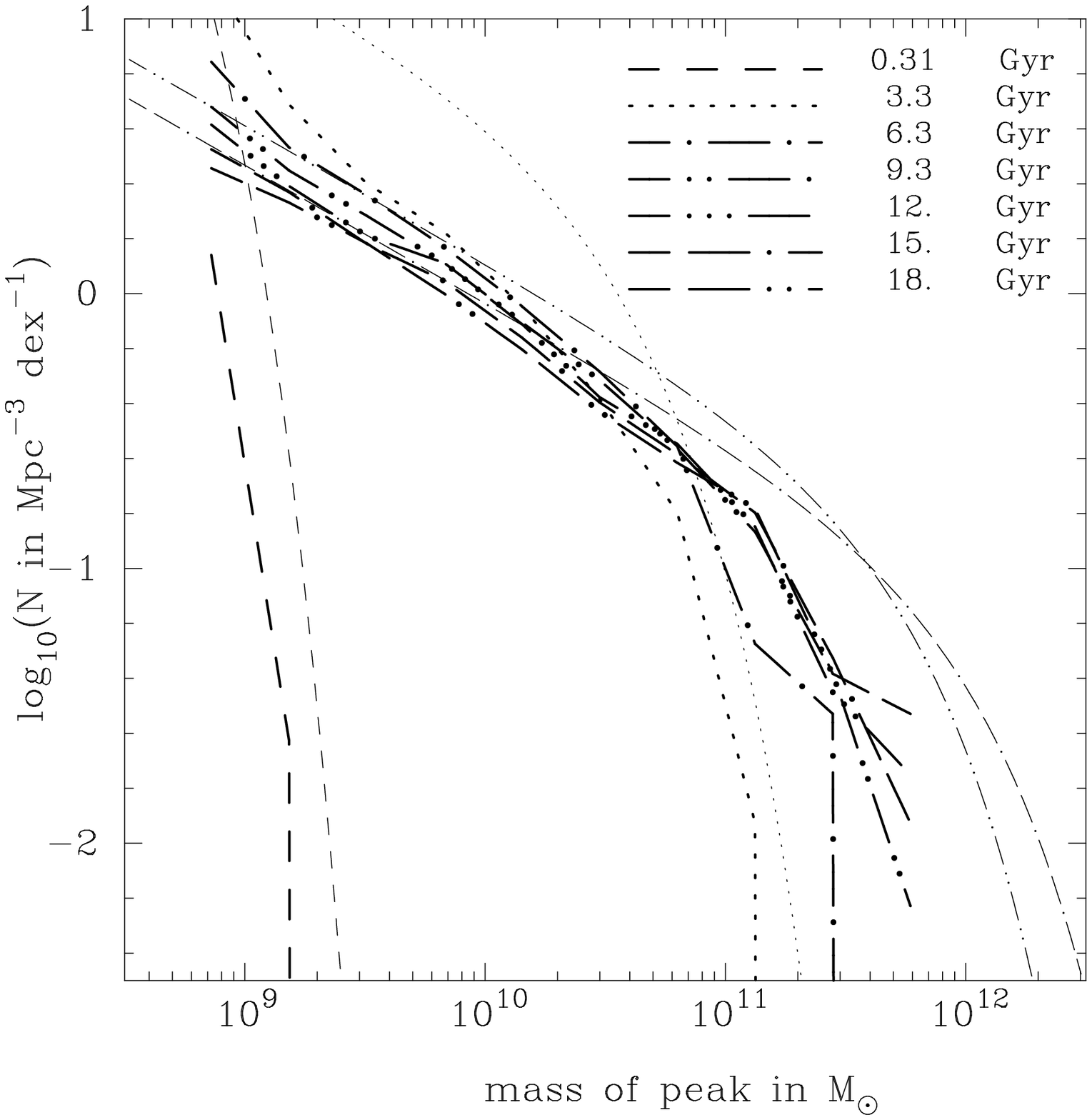"}}
}
::::
\subm \vspace{7cm} ::::
\caption[Mass functions for $n=0 $ model for $\protect\rthresh =1000.$]{ 
\label{f-mfn0br1000} Mass functions for $n=0 $ model for 
$\protect\rthresh =1000.$ Curves are as for Fig.~\ref{f-mfn-2br5} 
except that $\delta_{c0} = 5 \protect\dvir$ in the PS formula  
for the $1\.5$~Gyr time step 
and $\delta_{c0} = 2 \protect\dvir$ for the $3\.3$~Gyr time step.}
\end{figure}

} 

\def\fcorr{
\begin{figure}
\centering 
\nice \centreline{\epsfxsize=8cm
\zzz{   \epsfbox[25 200 578 534]{"`gunzip -c \datadir/corr.n-2b.r5.ps.gz"}}
 {  \epsfbox[25 200 578 534]{"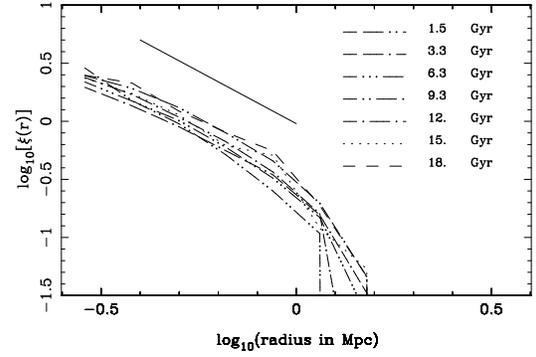"}}
}
::::
\subm \vspace{7cm} ::::
\caption[Spatial two-point autocorrelation functions of
density haloes (haloes) in $n=-2,\, \protect\rthresh =5$ model.]{ 
\label{f-corr.n0br5} Spatial two-point autocorrelation functions of
haloes in 
$n=-2,\, \protect\rthresh =5$ model, shown as $\log_{10}[\xi(r)]$
against $\log_{10}(r),$ where $r$ is the comoving halo pair
separation in Mpc. 
A solid line of slope $-\gamma=-1\.8$ is shown for comparison.}
\end{figure}

\begin{figure}
\centering 
\nice \centreline{\epsfxsize=8cm
\zzz{   \epsfbox[25 48 578 564]{"`gunzip -c \datadir/xi0evol.ps.gz"}}
  { \epsfbox[25 48 578 564]{"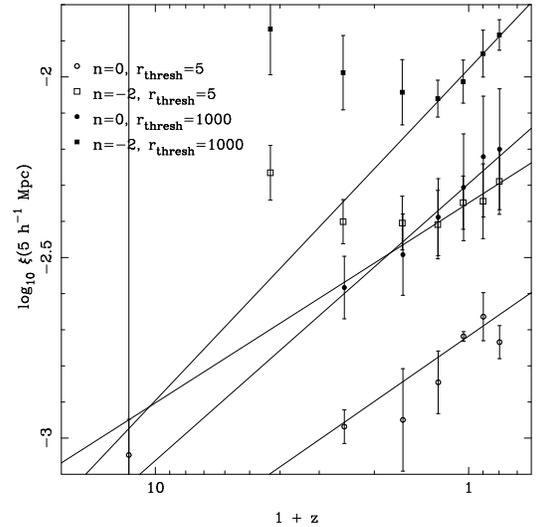"}}
}
::::
\subm \vspace{7cm} ::::
\caption[Evolution of spatial two-point autocorrelation function.]
{ \label{f-xi0evol} Evolution of the 
spatial two-point autocorrelation function, 
$\xi_0 \equiv \xi(5$~h$^{-1}$Mpc$, z)$ vs 
redshift ($1+z$).
(The highest redshift correlation function for $n=0,\, 
\protect\rthresh =1000$ is
too noisy to deduce a $\xi_0$ value.) 
Also shown are lines fitted to all but the highest $z$ point
for $n=0$ and to the four points with lowest $z$ for $n=-2$. }
\end{figure}
}

\def\finterp{
\begin{figure}
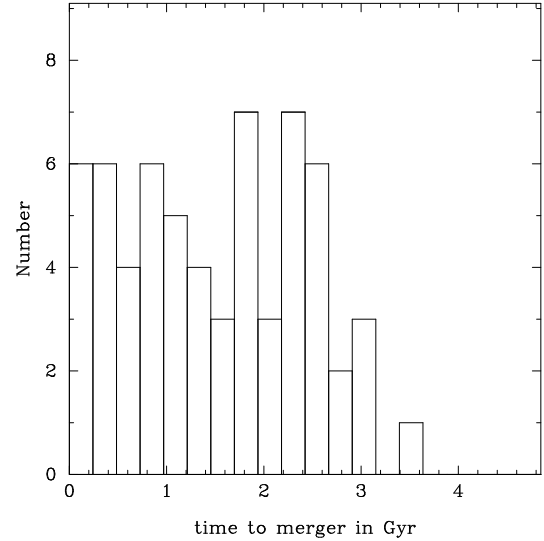

\centering 
\nice \centreline{\epsfxsize=7cm
\zzz{   \epsfbox[56 41 458 451]{"`gunzip -c \datadir/interp.ps.gz"}}
  { \epsfbox[56 41 458 451]{"interp.ps"}}
}
::::
\subm \vspace{7cm} ::::
\caption[Interpolation]{ 
\label{f-interp} Method by which time steps could be interpolated.
Histogram of time of radial infall of 
haloes (at $t=9\.3$~Gyr for $n=-2,$ $\protect\rthresh=1000$)
considered as point particles which fall into isothermal potentials
of the combined masses (at $t=12\.3$~Gyr) of the multiple merger 
products.}
\end{figure}
}   


\def\fsfr{
\begin{figure}
\centering 
\nice \centreline{\epsfxsize=7cm
 \zzz{  
\epsfbox[25 18 578 574]{"`gunzip -c \datadir/n-2b.r5.eb.sfr.ps.gz"}}
{  \epsfbox[25 18 578 574]{"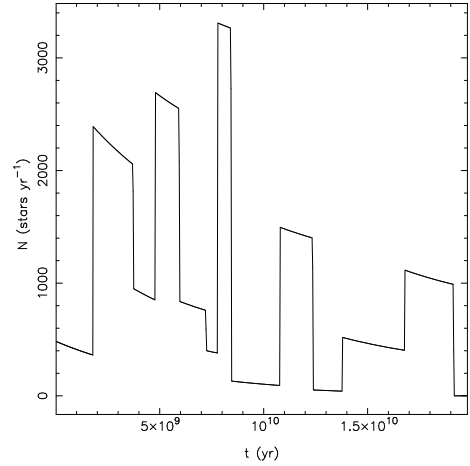"}}
}
::::
\subm \vspace{7cm} ::::
\caption[Example star formation rate (SFR).]{ \label{f-n0b.r5n6.eb.sfr} 
Star formation rate (SFR) history 
for the most massive galaxy at the final time step in the 
$n=-2,\, \protect\rthresh =5,$ exponential+burst model, $\mu=0\.15.$ 
The SFR in stars yr$^{-1}$ is plotted against time since the formation 
of the first progenitor of the galaxy.}
\end{figure}
} 

\def\flfevsb{ 
\begin{figure}
\centering 
\nice \centreline{\epsfxsize=7cm
\zzz{ \epsfbox[25 28 578 554]{"`gunzip -c \datadir/lf.n-2br5.eb.15.J.ps.gz"}}
 { \epsfbox[25 28 578 554]{"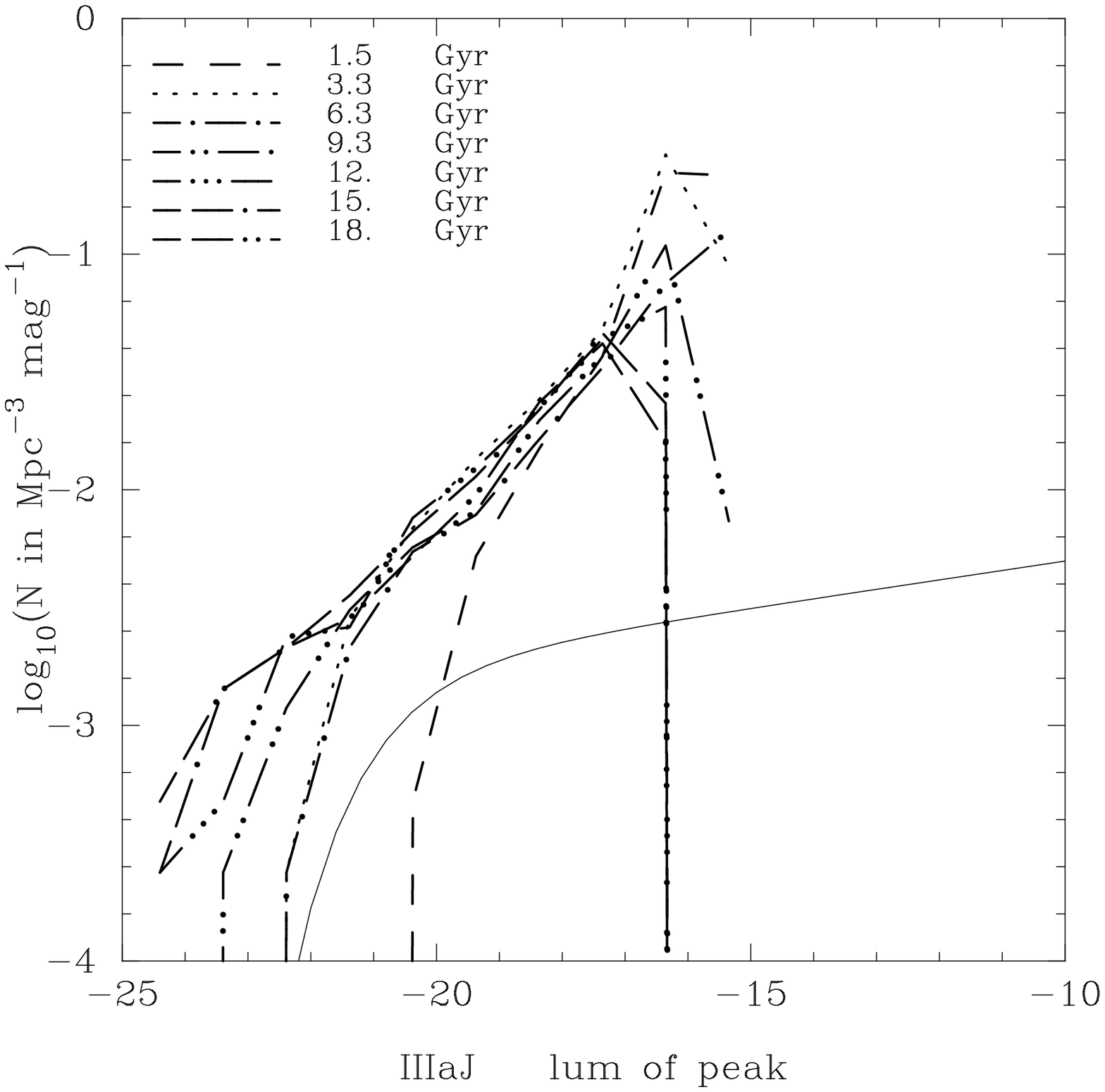"}}
}
::::
\subm \vspace{7cm} ::::
\caption[IIIaJ luminosity
 functions for $n=-2,\, \protect\rthresh =5$ model, exponential decay +burst, 
Bruzual's SFR index $\mu=0\.15.$  ]{     \label{f-lfn-2b.eb} 
IIIaJ luminosity functions for the 
$n=-2,\, \protect\rthresh =5$ model, exponentially decaying plus burst SFR, 
Bruzual's SFR index $\mu=0\.15.$ The model curves are of the line styles 
indicated, labelled by the beginning of each time stage interval, and 
compared with a Schechter function with locally estimated parameters 
(solid line, Eq.~\ref{e-schech}). 
Luminosities are expressed in absolute
magnitudes, $\MM\IIIaJ,$ in the observer's frame, i.e., 
K-corrected (\protect\cite{Wirtz}~1918);
densities are in $\log_{10}$(N~Mpc$^{-3}$~mag$^{-1}$) 	
in comoving coordinates. 
{\ \ \ } The following plots show the same quantities for different 
$n$ and $\protect\rthresh$. }
\end{figure}

\begin{figure}
\centering 
\nice \centreline{\epsfxsize=7cm
\zzz{ \epsfbox[25 28 578 554]{"`gunzip -c \datadir/lf.n-2br5.e.15.J.ps.gz"}}
{\epsfbox[25 28 578 554]{"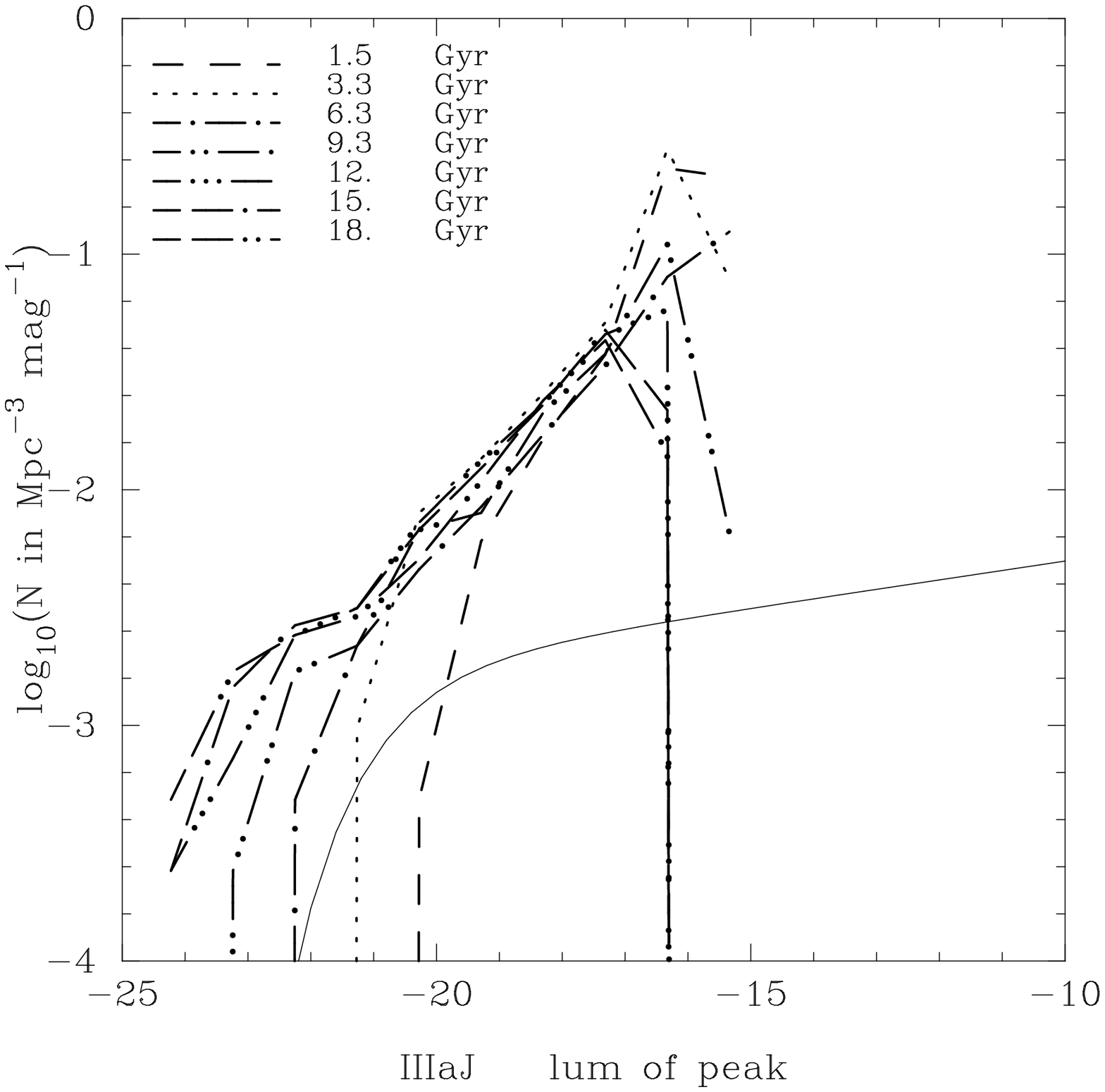"}}
}
::::
\subm \vspace{7cm} ::::
\caption[IIIaJ luminosity
 functions for $n=-2,\, \protect\rthresh =5$ model, (exponential decay only SFR),
Bruzual's $\mu=0\.15.$]{ 
\label{f-lfn-2b.e} IIIaJ luminosity
 functions for $n=-2,\, \protect\rthresh =5$ model, (exponential decay only SFR),
Bruzual's $\mu=0\.15.$ }
\end{figure}

\begin{figure}
\centering 
\nice \centreline{\epsfxsize=7cm
\zzz{   \epsfbox[25 28 578 554]{"`gunzip -c \datadir/lf.n-2br5.b.J.ps.gz"}}
{   \epsfbox[25 28 578 554]{"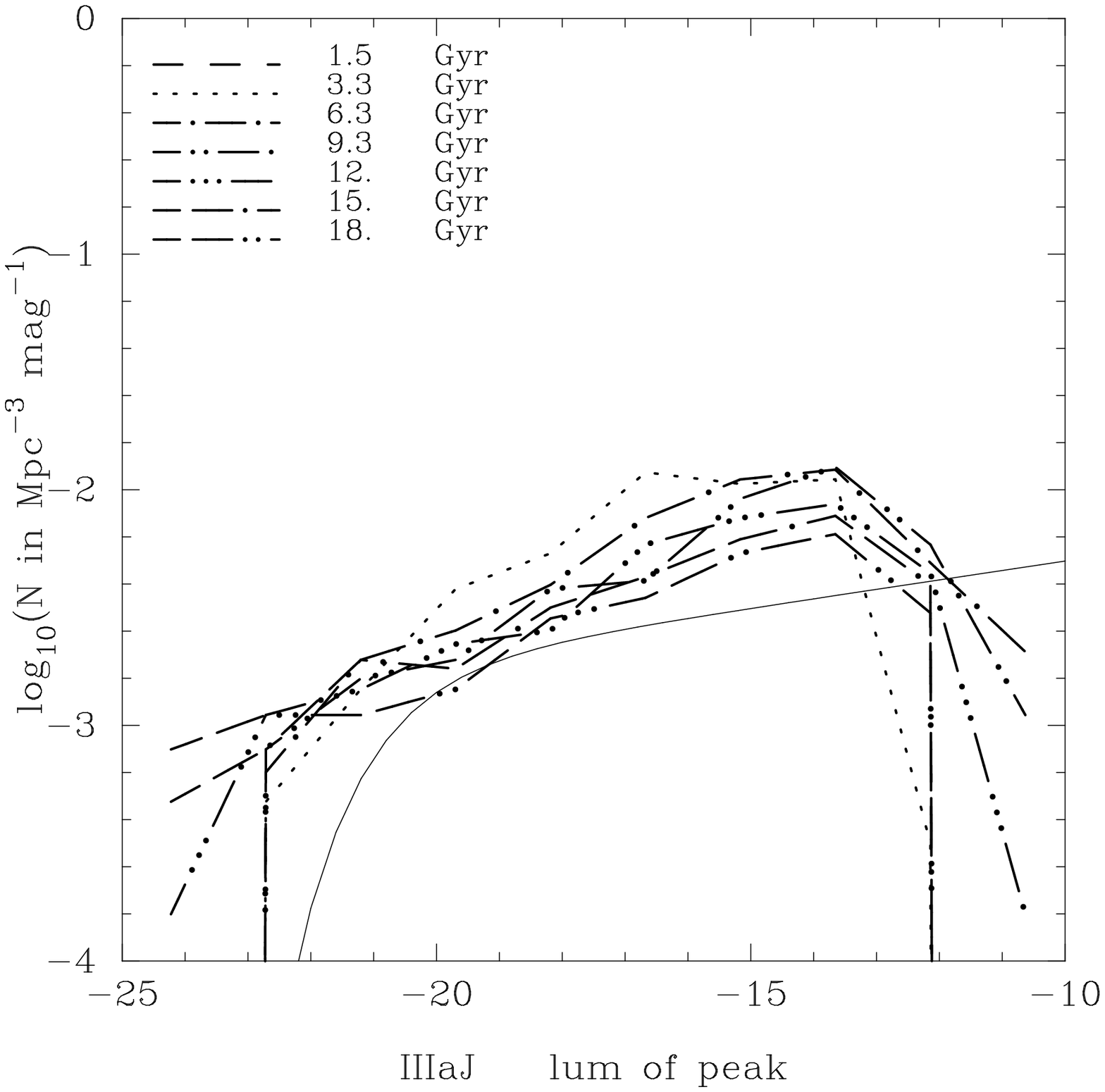"}}
}
::::
\subm \vspace{7cm} ::::
\caption[IIIaJ luminosity
 functions for $n=-2,\, \protect\rthresh =5$ model, burst-only SFR.  ]{ 
\label{f-lfn-2br5.b} IIIaJ luminosity
functions for $n=-2,\, \protect\rthresh =5$ model, burst-only SFR. 
In this and the following plots of the luminosity function for burst-only models,
the luminosity functions at the first time stage are missing. 
This is a property of
the model: stars are only formed when mergers occur; this first occurs at the
second time stage.}
\end{figure}
} 

\def\flfnrthr{ 
\begin{figure}
\centering 
\nice \centreline{\epsfxsize=7cm
\zzz{ \epsfbox[25 28 578 554]{"`gunzip -c \datadir/lf.n0br5.b.J.ps.gz"}}
{  \epsfbox[25 28 578 554]{"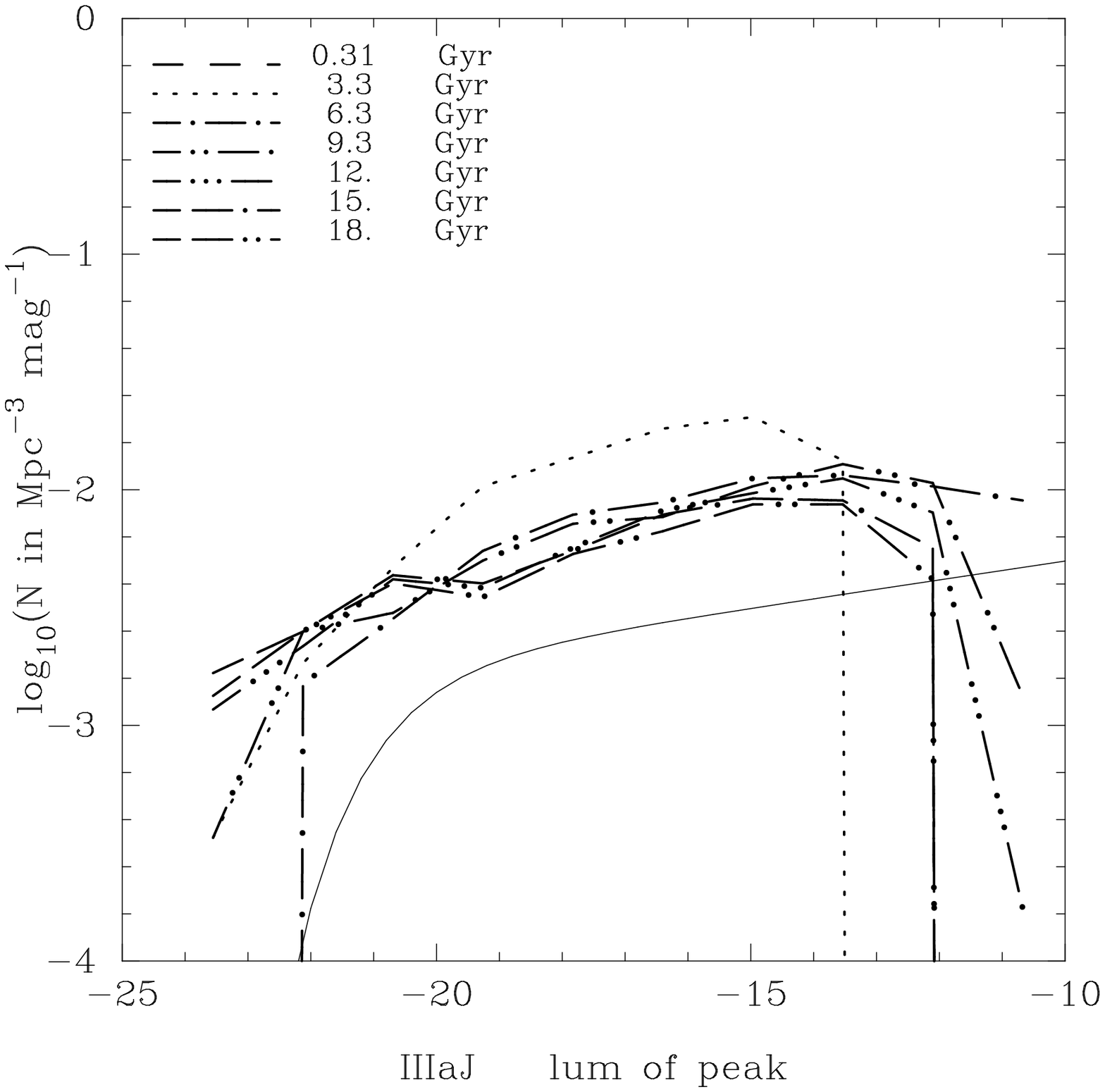"}}
}
::::
\subm \vspace{7cm} ::::
\caption[IIIaJ luminosity
 functions for $n=0,\, \protect\rthresh =5$ model, burst-only SFR.  ]{ 
\label{f-lfn0br5.b} IIIaJ luminosity
 functions for $n=0,\, \protect\rthresh =5$ model, burst-only SFR. }
\end{figure}

\begin{figure}
\centering 
\nice \centreline{\epsfxsize=7cm
\zzz{\epsfbox[25 28 578 554]{"`gunzip -c \datadir/lf.n-2br1000.b.J.ps.gz"}}
{\epsfbox[25 28 578 554]{"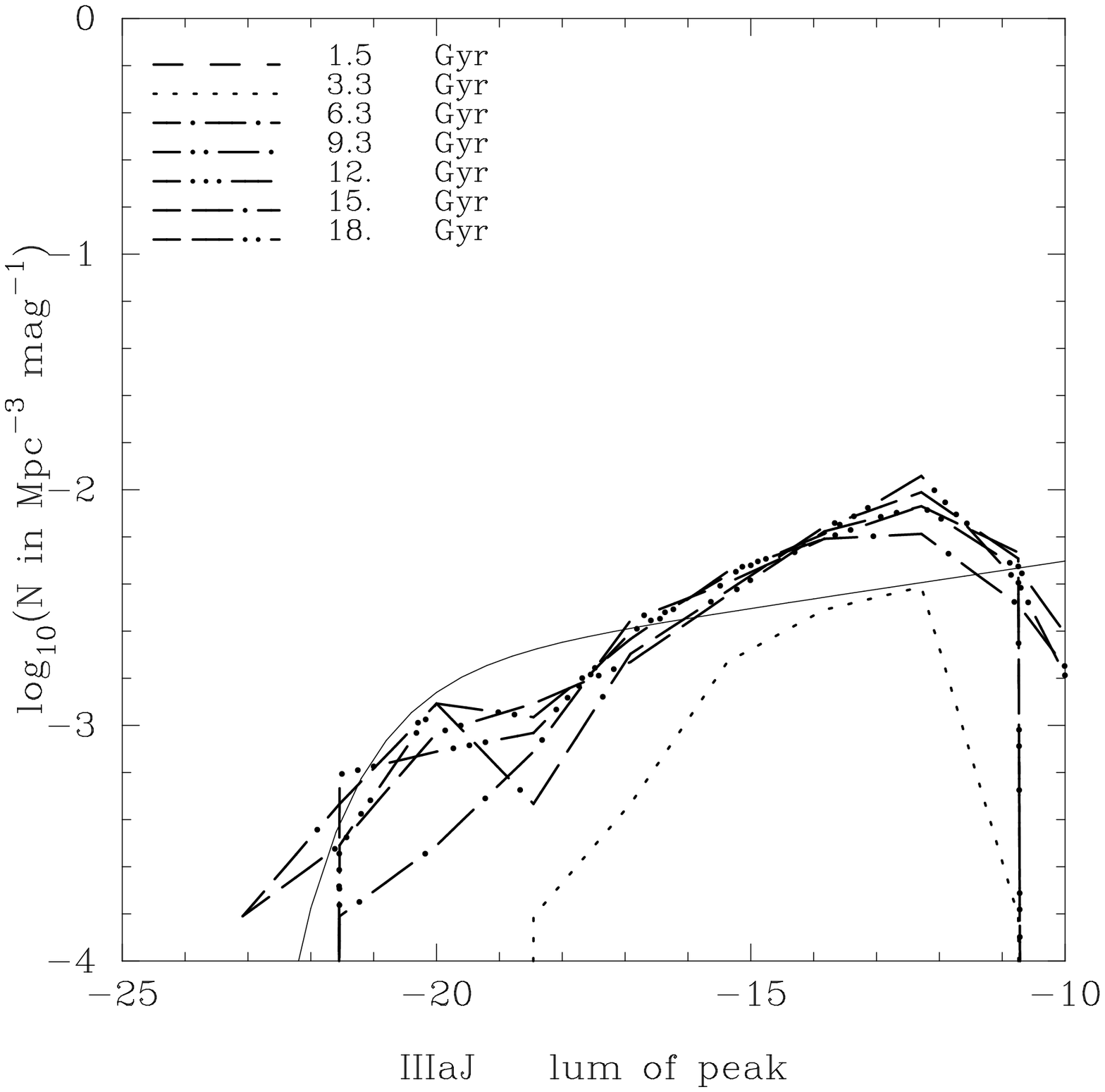"}}
}
::::
\subm \vspace{7cm} ::::
\caption[IIIaJ luminosity
 functions for $n=-2,\, \protect\rthresh =1000$ model, burst-only SFR.  ]{ 
\label{f-lfn-2br1000.b} IIIaJ luminosity
 functions for $n=-2,\, \protect\rthresh =1000$ model, burst-only SFR. }
\end{figure}

\begin{figure}
\centering 
\nice \centreline{\epsfxsize=7cm
\zzz{ \epsfbox[25 18 578 574]{"`gunzip -c \datadir/lf.n0br1000.b.J.ps.gz"}}
{\epsfbox[25 18 578 574]{"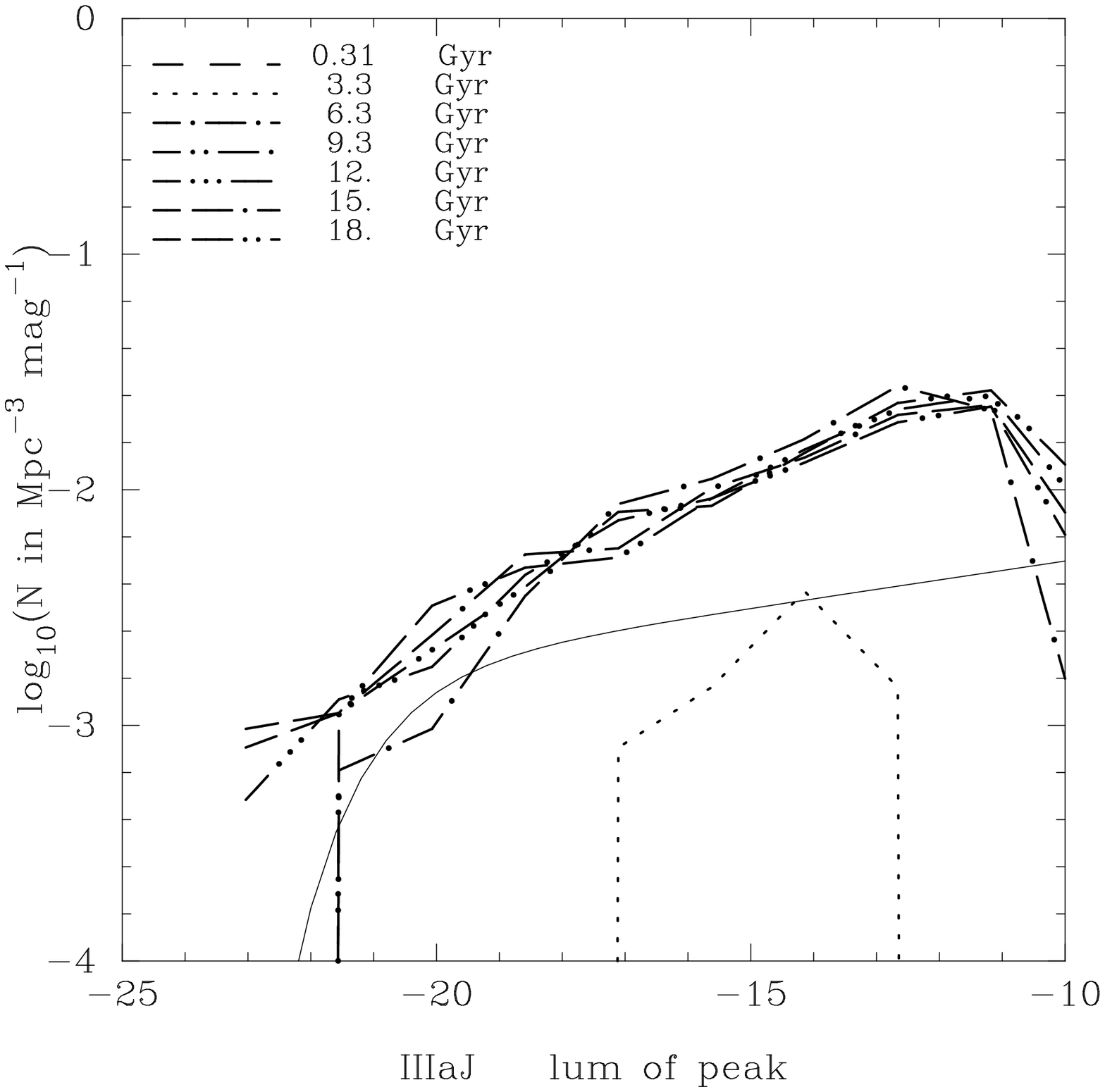"}}
}
::::
\subm \vspace{7cm} ::::
\caption[IIIaJ luminosity
 functions for $n=0,\, \protect\rthresh =1000$ model, burst-only SFR.  ]{ 
\label{f-lfn0br1000.b} IIIaJ luminosity
 functions for $n=0,\, \protect\rthresh =1000$ model, burst-only SFR. }
\end{figure}

}  

\def\fmlratio{
\begin{figure}
\centering 
\nice \centreline{\epsfxsize=7cm
\zzz{ \epsfbox[25 18 578 574]{"`gunzip -c \datadir/mln-2br5.eb.15.J.ps.gz"}}
{\epsfbox[25 18 578 574]{"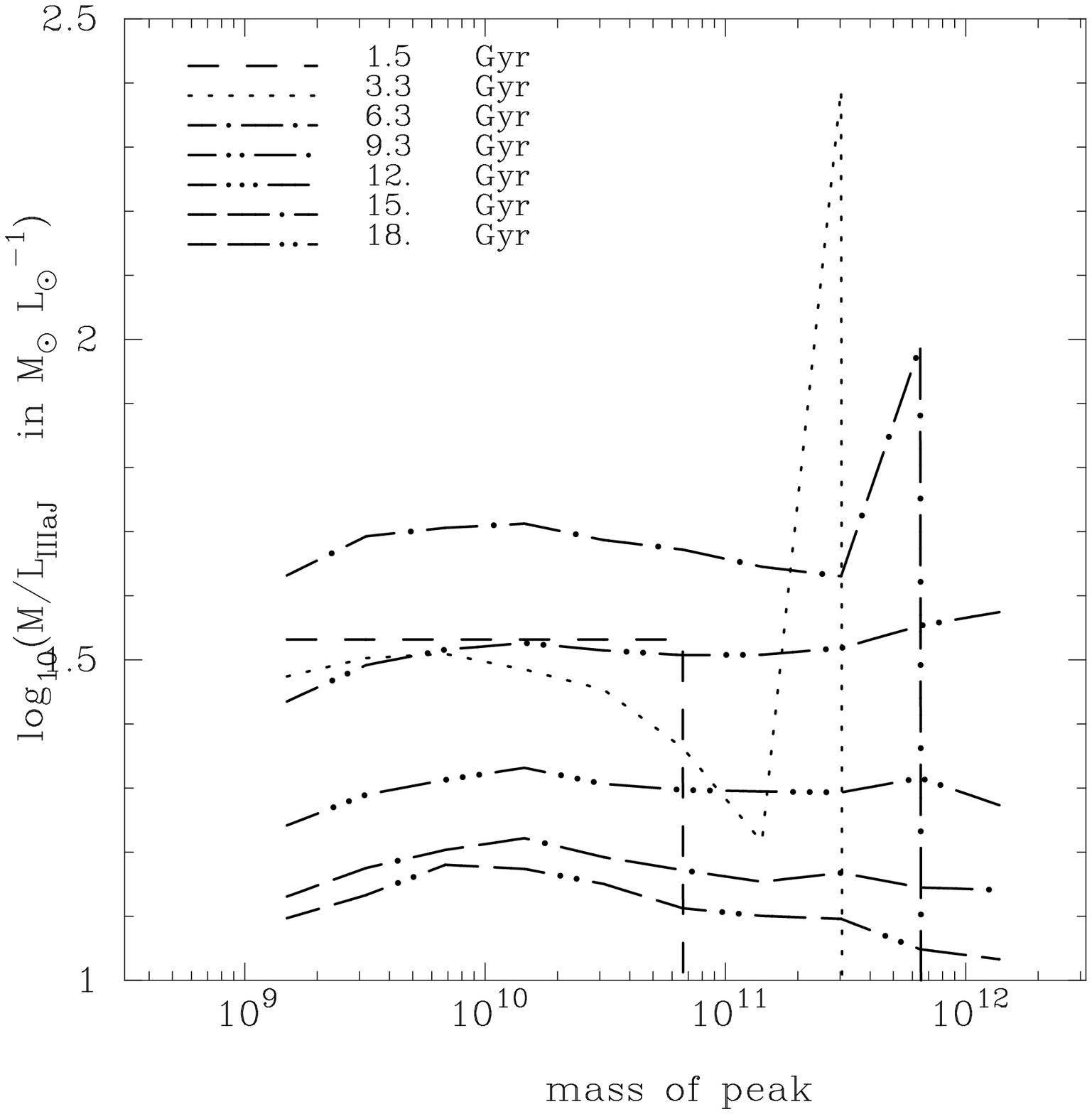"}}
}
::::
\subm \vspace{7cm} ::::
\caption[$\protect\MMlum/L\protect\IIIaJ$ values for 
$n=-2,\, \protect\rthresh =5$ exp+burst model.]{ 
\label{f-mln0b.eb} $\MMlum/L\IIIaJ$ 
for rest frame values of $L\IIIaJ$ for 
a run on the $n=-2,\, \protect\rthresh =5$ model merging history 
with exponential and burst evolution turned on and Bruzual's 
SFR parameter $\mu=0\.15.$  The masses detected at the time stage $t_i$
and the luminosities resulting at the end of the interval $[t_i,t_{i+1})$ are
used to obtain  $\MMlum/L\IIIaJ$ values labelled $t_i$ in this figure.}
\end{figure}
}   


\section{Introduction}  
    As quantitatively pioneered by those such as 
\cite{Hoyle}~(1953), \cite{Silk77}~(1977) and  \cite{ReesO78}~(1978) and more
recently noted by authors such as Frenk \etal\ (1995), hierarchical galaxy
formation models in $\Omega_0=1$ cold dark matter (CDM) universes 
typically combine assumptions on up to six distinct physical
processes: (1) the non-linear growth 
phase of matter density peaks (known as ``haloes''),
(2) cooling gas dynamics, (3) star formation, (4) star-to-gas
energy feedback, (5) stellar evolution, (6) galaxy mergers. In principle,
if there are more free parameters describing these processes
than independent observational galaxy statistics, 
then the observations should provide little constraint on
 galaxy formation ``recipes''.
Fortunately, the contrary is
presently the case for the ``semi-analytical {\em ab initio}'' models which 
make various 
analytical estimates of process (1), 
combine semi-empirical and simple scaling 
parametrisations to represent processes (2)-(4) and (6) and
use evolutionary stellar population synthesis for process (5). 
Since each of these models have problems explaining at least some 
of the observations
means, the models are better constrained than might
have been hoped for.

	These models can be considered to be semi-analytical because rather
than calculating what is possibly the most important process, 
the non-linear formation and merging history 
of collapsed objects [process (1)], via N-body simulations, various 
statistical analytical approximations are used.
The models of Lacey \etal\ (1993) use an approximation developed 
in \cite{LS91}~(1991) from the BBKS peaks formalism 
(\cite{BBKS}~1986),
Kauffmann et {al.} (\cite{KW93}~1993; \cite{KWG93}~1993)
use a probabilistic method (\cite{Bower91}~1991) 
based on the Press-Schecter formalism 
(e.g., \cite{PS74}~1974; White \& Frenk 1991) and 
excursion set mass function calculations (\cite{Bond91}~1991) while 
\cite{ColeGF94}~(1994) use a spatially quantised ``block'' method
described in \cite{ColeKais}~(1988). 

Further semi-analytical developments 
include those adding spatial auto-correlation information 
to a Press-Schechter formalism (\cite{MoWh96}~1996) or to the
``block'' model (\cite{RT96}~1996; \cite{Naga97}~1997) and a technique 
of separately treating global, weakly non-linear and local, strongly 
non-linear dynamics (the ``peak-patch'' formalism, \cite{BoMy96}~1996).

Each of the models which has been compared 
to observational statistics has difficulty in 
simultaneously explaining the flatness of the
present-day (surface-brightness limited) galaxy luminosity function
(e.g., Loveday \etal\ 1992), the steepness of the faint galaxy counts
(e.g., \cite{TSei88}~1988; \cite{Tys88}~1988), the shape of the moderately
faint galaxy ($B\ltapprox23$) spectroscopic redshift distributions 
(e.g., \cite{Coll90}~1990; \cite{Coll93}~1993), the Tully-Fisher
relation and the colour distributions of present-day galaxies, 
in a CDM $\Omega_0=1$ universe. Even though the \cite{ColeGF94}~(1994) model
is better than the previous models in allowing big $z=0$ galaxies to
be at least as red as higher $z$ galaxies, it 
shares the problem of the other models in lacking big red
ellipticals. It also shares with the \cite{KWG93} model the problem
that if the large number of small haloes predicted by CDM models at 
$z\approx0$
follow the IR Tully-Fisher relation (e.g., \cite{PTully92}~1992), 
then the slope of 
the faint end of the general galaxy luminosity function should be
steeper than that estimated locally (e.g., 
Loveday \etal\ 1992). Changing the cosmology in the 
\cite{ColeGF94} models 
(\cite{Heyl95}~1995: low $H_0$, low $\Omega_0$, non-zero $\lambda_0$ and
CHDM models) is insufficient to match the observations.
Another way of allowing these models to fit the
observations is to make a strong assumption for process (6)---to
suppose that
galaxies can merge as fast as galaxy haloes merge, or even faster---but 
simple present-day 
constraints on the products of the mergers 
(\cite{Dalc93}~1993) and the relative weakness of the faint galaxy 
angular auto-correlation function (\cite{RY93}~1993) strongly
restrict this possibility.

In order to avoid problems which may be due to the approximation of 
non-linear gravitational collapse and evolution by the 
semi-analytical techniques 
mentioned above, an alternative technique is to calculate both processes (1)
and (2) from first principles 
in numerical N-body simulations, folding in the other physical 
processes as simple scaling formulae or using stellar population synthesis
for (5). Several authors 
(e.g., \cite{Ev88} 1988; \cite{NavB91}~1991; 
\cite{CO92cdmhydro}~1992; \cite{Ume93}~1993; \cite{StM95}~1995) 
have 
experimented with these techniques, but resolution limits
on present-day computers make the results hard to interpret. For example,
\cite{WeinbHK97}~(1997) point out that although low resolution 
gravito-hydrodynamic simulations suggest that a UV photoionisation 
background can suppress galaxy formation (by heating the gas so that
it is unable to cool and form stars),  higher resolution simulations
show that this is a numerical artefact: the higher resolution 
simulations show little sensitivity to either the details of
photoionisation or star formation.

	In this article, rather than claiming a global ``recipe'' for galaxy
formation, our primary purpose is to concentrate on 
process (1) in a way complementary to that of other techniques. 
This is unlikely to be sufficient to solve all the observational conflicts.
On the contrary, this method should increase the ability of modellers 
to verify the extent to which model predictions are sensitive to the 
precision of modelling of gravity.

	The method presented here is 
to derive merging history trees of dark matter haloes 
directly from N-body simulations. 
Rather than just investigating virialised haloes for
a particular dark matter model (e.g., CDM), (a) both $n=-2$ and $n=0$
initial perturbation spectrum simulations 
(where $n$ is the index of the power spectrum)
are examined, and (b) since the halo-to-galaxy relationship may be
more complex than a simple one-to-one mapping, two significantly different
density thresholds are used for halo detection. 
This reveals the sensitivity of halo
merger history trees and halo statistics to these parameters. 
The N-body simulations used are presented in \S\ref{s-N-body}, 
the choice of a group-finding algorithm in \S\ref{s-peakalg} and 
the defining criterion and algorithm for calculating the merging history trees
in \S\ref{s-tree}.

	Properties of the haloes detected are discussed in \S\ref{s-halostats}.
In particular, the resulting merging history trees are presented in graphical
form in \S\ref{s-mhtrees}, enabling patterns of halo
merging calculated from fully non-linear simulations to be visualised directly.

	If processes (2)-(5) are simple enough, and if process (6), 
galaxy merging, corresponds in a one-to-one way with halo merging,
then these halo merger history trees would lead directly to galaxy
merger history trees. We therefore examine an example application
of the merger history trees by 
making minimal assumptions for processes (2)-(4), using stellar evolutionary
population synthesis for process (5), and for process (6), 
assuming maximal galaxy merging (every halo merger corresponds to
a galaxy merger). 
\S\ref{s-bursts} presents 
(6)+(3) merger-induced star formation and \S\ref{s-geps} explains how
process (5) is modelled.

In order for these processes to have an effect on
the luminosity function, an option is considered in which
each merger induces a burst of star formation, 
scaled according to the appropriate halo and gas masses
and the dynamical time scale. 
Apart from this star formation rate option, 
we do not explore parameter space for non-gravitational processes 
in this paper; we merely adopt simple observationally normalised 
scaling laws.
Resulting luminosity functions are presented in \S\ref{s-galstats}.

	Applications of N-body derived 
halo merger trees with more complex assumptions for processes (2)-(6)
are of course possible, and indeed to be welcomed. The galaxy formation
``recipe'' explored here is only one simple example. 

	Cosmological conventions adopted for this paper are a Hubble constant 
of $H_0=50 \,$km\,s$^{-1}$Mpc$^{-1}$, comoving units (at $t=t_0$) and 
an $\Omega_0=1\.0, \Lambda=0$ universe is assumed, 
except where otherwise specified.

\section{Halo Merger Histories (Gravity)}

\subsection{Method}

\subsubsection{N-body Models of Matter Density} \label{s-N-body} 
	The non-linear gravitational evolution of matter 
density is modelled by 
N-body cosmological simulations run by 
\cite{Warr92}~(1992). 
These simulations use a 128$^3$ initial 
particle mesh, of side length 10~Mpc. 
[The simulations analysed here 
are for power law initial perturbation spectra ($n=-2$ and $n=0$), so this is 
simply a default choice for the scaling of units. This default scaling 
is used hereafter except where otherwise specified.]
Particles are placed on this mesh, making a cube of $\sim 2\e{6}$
particles. 

An initial perturbation spectrum is imposed on this cube by Fourier transforming
the initial complex amplitudes from the perturbation spectrum and using 
the Zel'dovich
growing mode method (\cite{Warr92}~1992)
on this Fourier transform and the $128^3$ particle mesh.
The amplitude of the perturbation spectrum is chosen
such that linear perturbation growth implies that 
$(\delta M/M) (r=0\.5 \hMpc)=2\.0$ at $z=0,$ where $(\delta M/M)(r)$ 
is the r.m.s. value of the excess mass (over uniform
density) in spheres of radius $r$ (\cite{Warr92}~1992).
This choice ensures that the haloes which collapse are about the same size
for different values of $n,$ so that the dependence on $n$ of properties 
of halo dynamics---or merging histories---can easily be studied. 
The absolute normalisation of the spatial correlation function 
of the haloes cannot be directly interpreted in terms of observational
quantities. The relative amount of power on different scales 
(or slope of the spatial correlation function), and the halo detection
threshold, are the parameters which may affect the rates and ways in
which haloes merge with one another.

The initial cube of perturbed particles is trimmed to a sphere,
i.e., particles more than $5000\,$kpc from the centre of the 
cube are removed, resulting in a sphere of $\sim 1\.1\e{6}$ particles.

This is then evolved forward gravitationally via a tree-code 
(e.g., see \cite{BHut86}~1986), initially
with roughly logarithmic time steps up to $t=0\.3 \,$Gyr, after which equal 
time steps of $0\.03 \,$Gyr are used. Every hundredth time
step is stored on disk; these are the time steps available for
halo analysis (hereafter ``time stages''). A vacuum boundary condition 
is used and the softening parameter is 5~kpc (proper units).

\subsubsection{Group-Finding Algorithm} \label{s-peakalg} 
      The simulation data are searched for density peaks at each 
time step by an algorithm which 
uses the ``oct-tree'' method to find all overdense regions without 
overwhelming computer memory, followed by an iterative means of joining 
together contiguous overdense regions. 

	Alternative group-finding methods which could be used
include the ``friends-of-friends'' (FOF) algorithm 
(e.g., \cite{WDEF87}~1987), the algorithm used by \cite{Warr92}~(1992) 
or the DENMAX algorithm (\cite{GB94}~1994).

The FOF group-finder has the advantage of low memory requirements and
an obvious relation between the mean particle separation and the 
group-finding resolution, but has the disadvantages that if the link
parameter $l$ is too low, then low density haloes---or the low density 
envelopes of haloes---are missed, while if $l$ is higher, small but 
distinct haloes may be erroneously joined together as single objects.

\cite{Warr92}'s (1992) 
method, based on the accelerations of individual particles,
and the DENMAX algorithm, which includes a de-binding procedure to separate
haloes which are only temporarily close to one another, are both more
physically motivated than FOF. However, for a first investigation of 
the use of N-body generated merging history trees in galaxy formation 
models, the use of the simple method outlined below seems prudent. Since 
two different density detection thresholds are used, the implications 
of having either a low or a high fixed 
density threshold (which are similar to the cases of high or low $l$ 
respectively in FOF) can be seen. For further 
development, it would certainly be useful to consider
use of a more complex algorithm such as DENMAX.

	Details of the method are as follows.

      Conceptually, a cube concentric to the sphere of particles, 
having as side length the diameter of the sphere, is divided into
eight equally sized subcubes.
Any of these subcubes
containing more than one particle is itself subdivided into eight
subcubes. By not subdividing cubes 
with only one or zero particles, computer memory is not wasted on
analysing ``empty'' space.
The subdividing process is iterated to a depth of $\nlevels$
 levels below the original cube,
unless at some level 
all the cubes have one or zero particles in them, in which 
case subdividing stops (this would happen at $\nlevels=8$
for this $1\.1\e{6}$-particle model for a uniform particle
distribution).
The side length of the smallest cube is $174\,$kpc and $20 \,$kpc
for $\nlevels=6$ and $\nlevels=9$ 
respectively at $z=0.$ 

The ``primary'' list of density peaks is then simply the list of 
each cube at the deepest level (i.e., of size $2^{-\nlev_tiny}$ times the
simulation sphere diameter)  whose density is at or 
above $\rthresh$ times the mean density. The list of particles in
each of these peaks is recorded.

The results presented here are for $\rthresh =5$ and $\rthresh =1000.$ 
For a flat rotation curve of the Galaxy of $220 \k,$ 
the cumulative mass to a radius $r$ is $M(<r)\propto r,$
and the density is
$\rho(r)= 1\.2\e{7} \rho_c  r^{-2}$ for $H_0$ as above and $r$ in kpc.
So, detection densities 
of $\rthresh =5$ and $\rthresh =1000$
correspond to the total (baryonic plus nonbaryonic) matter
density at galactocentric distances of about 1500~kpc and 
110~kpc respectively. The latter is a reasonable
value for the halo radius, but the former is several times 
greater than the largest radii claimed for the halo of the Galaxy.

The key to a simple way to join together contiguous ``primary'' peaks
is to order the primary density peak list 
by mass, from largest to smallest, so that each ``secondary'' peak
can be created by starting from its nucleus (densest region) and successively
joining on regions of lower and lower density which are adjacent
to the region which has already been 
aggregated.\footnote{The idea to order the ``primary'' peaks by mass 
was inspired by the group-finding algorithm of \cite{Warr92}~(1992).}

The parameters used to decide on adjacency are 
the radius $r$ (from centre to outermost particle) of each secondary peak 
and an ``incremental radius'', 
$\rinc$, defined as $1\.1$ times half of the largest diagonal of the
small cube used in finding the primary density peaks. 
A primary peak is
considered adjacent to a secondary peak if its centre is within $r+\rinc$
of the secondary peak. 
The radius $r$ is re-evaluated each time a primary peak is 
joined to a secondary peak.
The nucleus of the first secondary peak is the
first (i.e., most massive) primary peak; each following secondary peak 
starts with the most massive primary peak not previously included in a 
secondary peak. 

The final secondary peak list, 
 corresponding to all separated regions inside isodensity contours
of $\rthresh$ times the mean density, is hereafter simply termed
a peak list, since the list of primary peaks is not of astrophysical
interest. 	
The members of this list are considered to be dark matter haloes.

\subsubsection{Creation of History Tree} \label{s-tree} 
      A peak (halo) merging history tree is obtained as follows. 

      Peak lists for a series of time stages of the model are obtained by 
the algorithm just described, each obtained with the same 
values of $\nlevels$ and $\rthresh .$ 
For each pair of successive output 
times, $t_i, t_{i+1},$ the peaks at the 
two times are compared. Two arrays, $a_i, a_{i+1},$ each 
with as many elements as the number of particles in the simulation, are 
created. For each element $j$ of array $a_i,$ 
the integer $k$ identifying
the peak that particle $j$ is a member of is assigned to $a_i(j)$, where this
is a peak according to the peak list for $t_i$. 
The array $a_{i+1}$ is evaluated in the same way using the peak list for
$t_{i+1}.$
If the particle is not a member
of any peak, a null value is assigned. A simultaneous sort 
is performed on arrays $a_i$ and $a_{i+1}$, 
permuting both in the same way in order that 
$a_{i+1}$ is a non-decreasing arithmetical sequence. 

The result is that with the new ordering of $a_i$ and $a_{i+1},$ 
(1) a peak $k$ at $t_{i+1}$ is represented by a contiguous list 
$a_{i+1}(j)$ to $a_{i+1}(j')$ (each containing the peak number $k$) 
and 
(2) $a_{i}(j)$ to $a_{i}(j')$ (for the same $j,$ $j'$) represent the
same particles and contain values $(k_1, k_2, ...)$ indicating the peaks
(at $t_i$) of which the particles were members.
In other words, the peak 
membership at $t_i$ of particles in a single peak at $t_{i+1}$ 
is listed in $a_{i}(j)$ to $a_{i}(j').$

      For any peak at $t_{i+1}$, if more than 50\% of the 
particles in any of the peaks at $t_i$  
are present in the peak at $t_{i+1}$, 
then the peak at $t_{i+1}$ is 
considered a ``descendant'' of 
the peak at $t_i$ and the peak at $t_i$ is 
a ``progenitor'' of the peak at $t_{i+1}.$
These links are represented by appropriate arrays. Due to the nature of this
algorithm, no peak can have more than one descendant, though it can certainly 
have more than one progenitor, which is allowed for by using what are, 
in effect, pointers. 

      By applying this comparison 
across each pair of successive times $t_i,t_{i+1}$, a 
representation of the peak merging history is obtained.

\subsection{Results} \label{s-halostats} 
	The method described above has been applied to
both an $n=0$ and
an $n=-2$ power law initial perturbation spectrum N-body model 
(labelled ``n0b'', ``n-2b'' by \cite{Warr92}~1992). 
Table~\ref{t-redsh} shows redshifts and cosmological times for the 
output timesteps
for these two models. 
The negative redshifts correspond to
future times according to the default time scaling. 
If the time unit chosen were different to the default, 
then these latter time stages could be moved into the past
or the present.

\begin{table}
\caption{\label{t-redsh} Parameters of Time Stages Used}
$$\begin{array}{c ccc}
\hline \multicolumn{1}{ c}{redshift} &
      \multicolumn{1}{c }{t (\mbox{\rm Gyr})} &
      \multicolumn{2}{ c }{time step} \\
      \hline  
\multicolumn{2}{ c }{} & \multicolumn{1}{ c }{n=0 } &
      \multicolumn{1}{c }{n=-2 } \\ 
      \cline{3-4}
       11\.2 & 0\.31 &  40  &(15)  \\ 
      3\.2 & 1\.51    & -    &55 \\
      1\.5 & 3\.3    &  140&  115\\
      0\.62 & 6\.3   &  240 & 215\\
      0\.25 &9\.3    &  340 & 315\\
       0\.039 & 12\.3& 440 &  415\\
       -0\.10 & 15\.3 & 540 &  515\\
       -0.203 & 18\.3& 640 &  615\\ \hline
\end{array}$$
\end{table}

 Haloes are detected in both of these models at thresholds of both
$\rthresh =5$ and $\rthresh =1000$
times the mean universe density
(for an $\Omega_0=1\.0, h=0\.5, \lambda_0=0$
universe.)
The former density threshold for detection should result in
haloes which have only just turned around from following the
smooth Hubble flow, while the latter should result in well-virialised haloes.
These thresholds should span most cases of interest.
Statistics of the haloes detected and comparison of these to 
results from other methods 
are presented in \S\ref{s-peaks} and \S\ref{s-cfother}; examples of halo
merging histories are displayed in \S\ref{s-mhtrees}; 
the spatial two-point autocorrelation functions of the haloes  
are discussed in \S\ref{s-nbcorr}; and a suggestion for further 
development by interpolation of merger times between time steps 
is outlined in \S\ref{s-interp}.

\subsubsection{Basic Halo Statistics} \label{s-peaks}
   The numbers of haloes are 
shown in Table~\ref{t-npks}. In the $n=-2$ model,
matter has not yet collapsed into haloes
at $t=0\.31 \,$Gyr, so the $t=1\.5 \,$Gyr time step was used instead. 

\begin{table}
\caption{\label{t-npks} Number of haloes found for the different power
spectra and detection thresholds.}
$$\begin{array}{ c cc cc }
\hline \multicolumn{1}{ c }{t (\mbox{\rm Gyr})} 
	& \multicolumn{2}{c }{\rthresh =5}
		& \multicolumn{2}{c }{\rthresh =1000} 
	\rthstrut \\
\cline{2-5}  
      & n=0  & n=-2  & n=0  & n=-2 \\ \hline
0\.3  & 3959 & -    & 238  & - \\
1\.5  & -    & 2086 & -    & 412 \\
3\.3  & 1539 & 1890 & 4214 & 1421 \\
6\.3  & 1053 & 1121 & 2695 & 1516 \\
9\.3  &  836 &  804 & 2121 & 1176 \\
12\.3 &  712 &  637 & 1891 & 923 \\
15\.3 &  629 &  492 & 1674 & 790 \\
18\.3 &  590 &  433 & 1524 & 672 \\ \hline
\end{array}$$

\end{table}

	The reality of these haloes is verified visually by rectilinear 
projections of a sample of the points for each halo,  
by radial particle count profiles  
and by an interactive program which plots a sampling
of all the points on a computer screen. The program offers   
optional colouring of a range of haloes in a 
colour different to both the particle
colour and the black background 
and allows real time rotation of the image in order to give
an intuitive feel for the three-dimensional shape of the data. 
Examples of haloes are given in 
Fig.~\ref{f-profile}, which shows the radial 
particle count 
profiles for four of the biggest haloes (by number) 
detected using 
$\rthresh=5$ in the final time stage of the $n=-2$ model.
The profiles are simply numbers of particles per spherical shell, so the
rapid decrease to zero shows that the density falls off 
faster than $r^2.$ Note that one profile has  
two prominent maxima, 
neither at $r=0$. This is because, as a closer visual investigation
of the haloes shows, a small proportion of the ``haloes''
are in fact fairly close binary haloes rather than single haloes. 
These binaries are usually quite uneven in size, so consideration of the 
halo as a single halo is still a good approximation. 

As noted above, 
use of a group-finder such as DENMAX (\cite{GB94}~1994) would be an 
elegant way to avoid the adoption of ``binary'' haloes as single haloes. 
Alternatively, 
a positional proximity based group-finder could separate bound overlapping
bound groups of particles by requiring an additional layer of analysis,
such as detection of ``haloes within haloes'', in which either $\rthresh$ 
or $l$ (for a friends-of-friends group-finder) would have to again be 
specified.

\fprof

As described in \S\ref{s-tree}, for each time stage, a halo at time $t_i$ 
is considered to merge into a halo at the following
time stage $t_{i+1}$ (or retain its identity) 
if and only if more than $50\%$ of the particles of the halo at time $t_i$ are
present in the halo at time $t_{i+1}.$
Table~\ref{t-percm} shows the means and standard deviations of the 
fraction of a halo at time $t_i$ present in a halo at time $t_{i+1}.$ 
By definition, these fractions are 
constrained to be greater than $50\%.$

\begin{table}
\caption{\label{t-percm} Statistics of fraction of halo at time stage listed
here contained in halo at following time stage.}$$\begin{array}{ c cc cc }
\hline \multicolumn{1}{ c }{t (\mbox{\rm Gyr})} 
	& \multicolumn{2}{c }{\rthresh =5}
		& \multicolumn{2}{c }{\rthresh =1000} 
             \rthstrut \\
\cline{2-5}  & n=0  & n=-2  & n=0  & n=-2 \\ \hline

0\.3 & 96\pm8\% & -              & 90\pm11\,\%  & - \\
1\.5 &  -       & 87\pm15\% &  -      & 76\pm12\,\% \\
3\.3 & 92\pm9\% & 84\pm14\% & 78\pm12\,\%  & 74\pm12\,\% \\
6\.3 & 92\pm9\% & 81\pm14\%  & 79\pm12\,\%  & 71\pm11\,\% \\
9\.3 & 93\pm8\% & 82\pm13\% & 81\pm12\,\%   & 72\pm10\,\% \\
12\.3 & 93\pm8\% & 82\pm13\% & 82\pm11\,\%   & 72\pm11\,\% \\
15\.3 & 94\pm8\% & 84\pm12\% & 84\pm11\,\%   & 74\pm10\,\% \\
18\.3 & -   & -              &               &           \\ \hline
\end{array}$$

\end{table}

\fmfns

   Figs~\ref{f-mfn-2br5}-\ref{f-mfn0br1000}
show the mass functions of these
haloes and comparable mass functions from various semi-analytical 
methods. Our N-body derived mass functions are interpreted in terms of merging 
rates in the following paragraphs and compared with 
other mass function calculations in \S\ref{s-cfother}.

The mass function figures suggest that with 
the detection threshold of $\rthresh =5,$ 
for either 
spectral index the overall halo merging rate
from  $t= 3\.3 \,$Gyr 
(i.e., for $z \approx 1\.5$), to the present  
is little more than about
$ 3 - 10$ for galaxies below about $10^{10} M_{\sun}.$ While this
merging goes on, the number of large
haloes in the largest bin in the $n=0 $ model increases somewhat until
the last time step. Depending on the average number of small haloes which 
merge into a single large one, the increase in the number of large haloes would
appear at first sight to be explained by the decrease in the number of smaller
ones, consistent with a merging ratio of about $3-10.$ Though the high mass
end of the $n=-2 $ mass function is noisy,
similar interpretation could be made.

	These plots show a significant dependence on detection threshold
and a weak dependence on $n.$
	For the haloes detected at 
$\rthresh =1000,$ the merging is much weaker than for $\rthresh =5.$ 
In a given simulation, objects detected at the higher threshold 
consist of the dense cores of the objects detected at the lower threshold.
Hence, a simple explanation for the weaker merging is that 
if the low density envelopes merge, the cores don't necessarily do so,
but if the cores merge, the large low density envelopes are almost 
certainly going to merge. 

	However, some simple statistics show that the merging history
is not as simple as an $N$:1 ratio
applying equally to all haloes.

	Table~\ref{t-nnodesc} shows the fraction of the haloes
 at each time stage that have no descendants, i.e., the 
fraction of the haloes for
which no more than $50\%$ of their particles appear in any single halo
at the following time stage. The fact that these are nonzero 
(from about $5\%$
for $n=0,\, \rthresh =5$ to $30\%-50\%$ for $n=-2,\, \rthresh =1000$) 
shows that many haloes are destroyed
in the sense that more than $50\%$ of their particles may have been
pulled into an ``atmosphere'' of a large halo at a density lower than the
threshold density or possibly thrown out of the halo 
or pulled into another halo. This means that the halo number density does
not only decrease by merging, it also decreases by this halo destruction.
For example, if the overall number ratio between two time stages is {4:1}, 
but one in four haloes 
terminates, then the underlying ratio of haloes actually merging is only
{3:1}. Of course, this distinction is dependent on the definition of halo
detection identity as described above.

\begin{table}
\caption{\label{t-nnodesc} Fraction of haloes which have no descendants at 
following time stage.}
$$\begin{array}{ c cc cc }
\hline \multicolumn{1}{ c }{t (\mbox{\rm Gyr})} 
	& \multicolumn{2}{c }{\rthresh =5}
		& \multicolumn{2}{c }{\rthresh =1000} \rthstrut \\
\cline{2-5}  & n=0  & n=-2  & n=0  & n=-2 \\ \hline

0\.3 & 5\% & -    &  11\%  & - \\
1\.5 & -   & 15\% & -      & 32\% \\
3\.3 & 8\% & 24\% & 32\%  & 40\% \\
6\.3 & 8\% & 30\% & 26\%  & 49\% \\
9\.3 & 7\% & 29\% & 23\%   & 46\% \\
12\.3 & 7\% & 24\% & 20\%   & 44\% \\
15\.3 & 4\% & 21\% & 15\%   & 36\% \\
18\.3 & -   & -    &   -    &   -   \\ \hline
\end{array}$$
\end{table}

	More direct statistics are those of the histories of the
haloes detected at the final time stage. The mean (and standard deviation) 
of the overall number of haloes which 
collapse to above the threshold density 
(either at the first time stage or at a later time stage) and end up in
a final halo is shown in Table~\ref{t-nances}.

\begin{table}
\caption{ \label{t-nances} Numbers of original haloes which end up in a halo
detected at the final time stage (mean$\pm$st.deviation).}
$$\begin{array}{ c cc }
\hline \rthresh \rthstrut  & n=0 & n=-2 \\ \hline
5   &  7\.4 \,\pm\, 20\.7  & 5\.0 \,\pm\, 16\.9 \\
1000 & 3\.2 \,\pm\,  6\.5  & 2\.6 \,\pm\,  6\.2 
 \\ \hline
\end{array}$$
\end{table} 

While these mean values are in the range $3-10$ estimated above, the 
standard deviations show that many final haloes come from as many as 20 or more
original haloes. In fact, the maximum number of haloes that any final halo
originates from is $233$ for the $n=0$ model and $259$ for the $n=-2$
model (for $\rthresh =5$). For $\rthresh =1000,$ the overall rate is lower,
and the maximum numbers of haloes per any final halo are $88$ and $95$ for 
$n=0$ and $n=-2$ respectively.

\begin{table}
\caption{ \label{t-rainfrac} Mass fraction of a final halo which comes from
matter directly accreted from the ``background'' (mean$\pm$st.deviation).}
$$\begin{array}{ c cc }
\hline  \rthresh \rthstrut  & n=0 & n=-2 \\ \hline
5   &  32 \,\pm\, 26  & 23 \,\pm\, 28 \\
1000 & 36 \,\pm\,  25  & 29 \,\pm\,  32
\\ \hline
\end{array}$$
\end{table} 

	As already suggested by the number of haloes which terminate, throwing
matter back out into the background,
the amount of matter which ``rains'' or accretes onto haloes directly rather
than first collapsing into smaller density haloes is non-negligible. 
Typically 30\% of a final halo's mass comes from matter directly 
accreted from the background, but this fraction varies widely between 
individual haloes. Moreover, these fractions are little dependent on 
$n$ and $\rthresh,$ as can be seen in the statistics listed in 
Table~\ref{t-rainfrac}.

\subsubsection{Comparison with Other Methods} \label{s-cfother}

What advantages and disadvantages does this method have relative to 
others, e.g., the BBKS peaks formalism method used in 
\cite{LGRS93}~(1993) or the peak-patch formalism of \cite{BoMy96}~(1996)?

The advantages are that the simplifications made in 
the semi-analytical methods are not made in the N-body simulation.
For example:  
\begin{list}{(\roman{enumi})}{\usecounter{enumi}}
\item Most of the semi-analytical methods assume spherically symmetric 
collapse (though \cite{BoMy96}~1996 also consider the collapse of homogeneous
ellipsoids). \cite{Warr92}~(1992) show that the haloes in the simulations 
analysed in this paper are in fact triaxial. 

\item Only the most recent (e.g., \cite{RT96}~1996) 
methods include spatial two-point 
auto-correlation function information. Again, this
is automatically included in the N-body derived merging history trees 
(see \S\ref{s-nbcorr} for halo correlation functions). The same applies 
for higher order correlation information.

\item Characteristic dynamics of halo mergers such as tidal tails and
particles thrown out into low-density ``atmospheres'' (\S\ref{s-peaks}) 
are modelled in the N-body simulation but 
not taken into account in semi-analytical models.
\end{list}

Disadvantages include scientific problems such as 
resolution limits and practical problems such as managing large data files 
which occupy large amounts of disk space. 

A first order illustration of the similarities and differences between 
this method and others is obtained by examining the mass function plots
(Figs.~\ref{f-mfn-2br5}-\ref{f-mfn0br1000}). 
The Press-Schechter (PS) formula 
derivation of \cite{LC93}~(1993), using their Eqs.(2.11) and 
(2.1)\protect\footnote{The factor 
of $2/20$ in Eq.(2.1) of \protect\cite{LC93}~(1993) should read $3/20$.}] 
provides a simple analytical description of both the shape and 
the time evolution of the mass functions. 
PS functions for two early and two later time 
steps are shown for comparison in each mass function plot, and the values
of $\delta_{c0}\equiv\delta_c(z=0)$ 
are increased (from the default $\delta_{c0}=\dvir \equiv 1\.686$) 
in some plots in order to show ways in which 
the PS formula could be used to interpret the mass functions. 

For the $n=-2$ mass functions, our results bracket the 
semi-analytical model results for CDM 
initial fluctuation spectra, which have roughly this slope on galaxy 
scales. 
Shown on these plots are \cite{LS91}'s~(1991) mass function 
(Fig.~1; collapsed peaks) and \cite{BoMy96}'s (1996, Fig.~12) mass functions. 
(The latter are calculated for clusters, so 
masses and $\sigma_8$ are rescaled according to a length scaling 
$L \rightarrow L/20$.)

The time evolution of the PS functions, both in normalisation and shape, 
is similar to that of our haloes. This is best seen in 
Fig.~\ref{f-mfn0br5}, where the decrease in normalisation in proportion
to $(1+z)$ [expected from Eq.(2.1) of \cite{LC93}~(1993)] 
is clearly seen. 

The absolute normalisation of the PS functions is only 
about $1-3$ times that of the $\rthresh=5$ mass functions. Since the
PS formula is not intended to model $\rthresh=5$ or $\rthresh=1000$ 
haloes, better agreement should not be expected.

Indeed, Fig.~\ref{f-mfn-2br1000} shows that the 
haloes detected at $\rthresh=1000$ have a considerably lower 
normalisation than that of the PS formula, but similar to that of
\cite{BoMy96}'s prediction (rescaled to galaxy scales) for 
ellipsoidal internal halo dynamics. Since the N-body derived haloes
analysed here have a variety of triaxialities, it is not surprising that
this provides the best agreement. However, as remarked upon by 
\cite{BoMy96}, their method may not be ideal for small haloes, since
the low mass end ``must find the nooks and the crannies'' left over 
from the analysis devoted to high mass haloes, so an N-body method with
sufficient resolution may be better for obtaining the low mass slope of
mass functions.

\cite{LS91}'s~(1991) peaks formalism based mass functions are somewhat steeper 
than the present-day PS mass functions and do not provide the optimal 
fit to our mass functions.

Although the early epoch high-mass ends of the N-body mass functions 
for $\rthresh=5$ (Figs~\ref{f-mfn-2br5}, \ref{f-mfn0br5}) are
similar to the PS cutoffs (for $\delta_{c0}=\dvir$), the 
corresponding cuvres for the high density haloes ($\rthresh=1000$) can only be
matched to PS curves by increasing $\delta_{c0}$ 
(Figs~\ref{f-mfn-2br1000}, \ref{f-mfn0br1000}). Since 
$\delta_{c0}$ signifies a critical overdensity 
(according to a linear growth rate), 
it is not surprising that the high density halo high-mass cutoffs can be fit.
(However, this would not enable the PS formulae to 
fit the later time steps, since the normalisation increases in proportion 
to $\delta_{c0},$ which would worsen the disagreement for the high density
haloes relative to the PS prediction.)

\fmratios	

Of course, what is potentially
most interesting for explaining the flatness in observed 
(high surface brightness galaxy) luminosity 
functions (e.g., \cite{EEP88}~1988) is the agreement in the 
N-body derived ($\rthresh=5$) 
and PS slopes for $n=0.$ 
(The PS function has the 
form $\log_{10} dn/dM = [(n-3)/6] \log_{10}M + $const, 
for haloes of mass $M$).
An initial 
fluctuation spectrum slope of $n=0$ implies a less steep low mass (faint)
end of the mass (luminosity) functions than for $n=-2.$

More subtle merging properties, which can be compared with 
\cite{KW93}'s~(1993) PS derived Monte Carlo simulations, 
are the 
mass ratios of haloes and their most massive progenitors. The 
statistics of these ratios 
can be used to constrain the relation between galaxies 
and haloes via implications for the survival of disk galaxies. 

Fig.~\ref{f-rntn1_r5} 
shows that the mass ratios of merging haloes found
in \cite{KW93}'s~(1993) PS derived Monte Carlo simulations are 
similar to those found in our analysis for haloes detected 
at $\rthresh=5,$ as for the mass functions. Since the conditions of the 
simulations are somewhat different 
(PS vs N-body, CDM vs $n=-2$), 
the agreement suggests that these ratios are quite
robust with respect to different modelling techniques. 

Comparison with Fig.~\ref{f-rntn1_r1000} shows that at any given
time, the fraction of a high density halo which has already collapsed 
or accreted
is statistically lower than for low density haloes.

\mhplots	

\subsubsection{Merging History Trees}  \label{s-mhtrees}
Merging history tree plots 
(Figs~\ref{f-mhn-2b.r5.1_5}--\ref{f-mhn-2b.r5.400_410})
are obtained by choosing a range of haloes at the final time
stage and tracing back all the progenitors of these haloes. 
The line segments joining the circles are the key feature of the plots. 
These indicate that the halo at the earlier time is considered to merge into 
(or be ``identical'' to) 
the halo at the later time according to the 50\% criterion 
explained in \S\ref{s-tree}.
The aim of the plots is to show connectivity over time. So, the
horizontal axis is designed to separate the haloes according to their
future merging activity. It  
doesn't directly indicate space positions, although there
should be some correlation between how close two haloes are in the plot and
how close they are in space (since haloes need to be close 
in order to merge together later on).

      Much information on the merging process is represented in these tree
plots.  However, they do not show the entire complexity of 
the merging process (or ``graph'' in mathematical terminology). 
Since the halo merger history trees presented here 
start with a range of final haloes and trace backwards, 
haloes which have no descendant at the final time
stage are not shown.
As summarised in Table~\ref{t-nnodesc}, a significant fraction 
of the haloes at a given time stage have no descendants at the 
following time stage. A simple explanation is that a large
fraction of a halo can evaporate in the merging process and contribute 
less than 50\% of the mass of the final halo, 
in which case the merging/identity criterion adopted fails to find 
a descendant.
Typically, about $25\% \pm 25\%$ of
a halo evaporates (e.g., in tails) or forms low-density ``atmospheres''
in N-body simulations of $N\sim 2$ 
interacting galaxies (Quinn, 1992). These tails or
atmospheres are likely to fall below the density detection threshold.

Figs~\ref{f-mhn-2b.r5.1_5} and \ref{f-mhn-2b.r1000.1_5} 
show that merging ratios of up to around 10:1 occur for many of the most massive 
haloes at low redshifts, while as indicated by the maximum number of original
haloes for any final halo, the merging ratios between early
time stages can be much higher, as high as a few hundred to one 
in several cases for $\rthresh =5.$ 

The difference in detection thresholds 
appears clearly in the difference between the early time stages of 
Figs~\ref{f-mhn-2b.r5.1_5} and \ref{f-mhn-2b.r1000.1_5}. At the nominal redshift 
$z=11$, many low mass haloes have reached the turnaround density and are thus
detected above $\rthresh=5,$ after which they merge rapidly. In contrast, 
haloes detected above $\rthresh=1000$ are mostly detected at 
$z=1\.5$; the few which form earlier do not merge in ratios as high as for
the $\rthresh=5$ haloes. Depending on the assumptions one wishes to make
regarding the relationship between galaxies and haloes, either of these
limiting cases could be interesting. 

For merging history trees of the haloes having lower masses at the final
time stage (Figs~\ref{f-mhn-2b.r5.50_60} and \ref{f-mhn-2b.r5.400_410}), 
very little merging occurs apart from the earliest few time stages.  
Indeed, Fig.~\ref{f-mhn-2b.r5.400_410} shows that many of the smallest haloes 
either have only recently 
collapsed or are unmerged objects which have formed well after the first
time stage.  
	This can be generalised by stating that 
the larger a galaxy halo is, the more original haloes it is likely to have
been created from, and at any time in general, the more massive a galaxy halo
is, the more haloes are likely to be merging into it. (This effect can also
be seen in Figs~\ref{f-rntn1_r5} and \ref{f-rntn1_r1000}.)

   That the lowest mass peaks may form fairly recently or may
form early yet undergo no merging at all is to be expected 
in a ``hierarchical galaxy formation'' scenario. Bottom-up
gravitational collapse 
does not only imply that small haloes form first and successively merge
to form more and more massive haloes, but also that 
some low mass haloes (e.g., from
low amplitude short length-scale perturbations, which must exist if the
initial perturbation amplitudes are part of a zero-mean Gaussian distribution)
continue to form at late epochs. 

	The recently forming small haloes in our 
halo merger trees could be used to model the 
(observed) existence of low mass, young, low redshift
galaxies, such as the dwarf Spheroidals. An interesting case is 
 the SBS\-~0335-052(W) pair of dwarf galaxies at
a redshift of $z=0\.014$, which have a (stellar) age of not more than $10^8$~yr
and have a metallicity (O/H) around 1/40 of the solar value 
(\cite{Lipo97}~1997).  
These two dwarfs are 
extremely hard to understand as anything other than young, low redshift,
low mass galaxies. For such galaxies to provide a constraint on galaxy 
formation models, precise estimations of number densities and age 
distributions would be useful---though obviously difficult to obtain without 
strong observational selection effects.

\fcorr

\subsubsection{Halo Correlation Functions}  \label{s-nbcorr}
Another important statistic of the haloes 
is their spatial two-point
auto-correlation function, $\xi(r).$ The natural inclusion of $\xi$ in
N-body simulations is something not usually present in semi-analytical
galaxy formation modelling. Indeed, \cite{YNG96}'s (1996)
application of \cite{Jdzk}'s (1995) approach to the ``cloud-in-cloud'' 
problem of the Press-Schechter formalism (\cite{PS74}~1974) shows that
the mass function of the latter suffers significantly from the lack of
inclusion of $\xi.$ \cite{Naga97}~(1997) confirm this in semi-analytical
merging history tree simulations using 
\cite{RT96}'s (1996) 
modification of the Block model (\cite{ColeKais}~1988).

  Fig.~\ref{f-corr.n0br5} shows the correlation functions for the
haloes detected for $n=-2,\, \rthresh =5.$ These functions have 
slopes $-\gamma\approx -1\.8,$ where 
\begin{equation}
\xi= (r_0/r)^{\gamma} (1+z)^{-(3+\epsilon-\gamma)},
\label{e-defneps}
\end{equation}
$r$ and $r_0$ are expressed in comoving coordinates, 
and $\epsilon$ parametrises the growth rate of $\xi_0(z)\equiv\xi(r_0,z)$, 
deduced either from observation or theory (\cite{GroP77}~1977).

 The slopes of $\xi$ for the other combinations of $n$ and $\rthresh$ 
are similar, so  
are consistent with that observed for galaxies, i.e., 
$1\.7 \ltapprox \gamma \ltapprox 1\.8$ (e.g., \cite{Bible}~1980; 
\cite{DavP83}~1983; \cite{Love92}~1992). Some observations indicate that 
$\gamma$
evolves with time (\cite{IP95}~1995), but others do not detect this 
(\cite{HudLil}~1996). N-body simulation derived correlation 
function slopes vary around 
the same values, but are sensitive to whether the correlation
is that of particles (identified as galaxies), of haloes or of mass-density
(\cite{DEFW85}~1985; \cite{EFWD88}~1988;   
\cite{SutoS91}~1991; \cite{SugiS91}~1991; \cite{Melo92}~1992).

The haloes of our N-body derived merging history trees therefore have realistic 
correlation slopes and do not suffer from the lack 
of a correlation function of the traditional semi-analytical methods.

Another interesting property of the correlation function is the growth 
rate of $\xi_0,$ parametrised by $\epsilon.$ This is shown 
in Fig.~\ref{f-xi0evol}. 
Since the simulations used here are not normalised 
conventionally, the values of $\xi_0(z=0)$\footnote{These 
could be reinterpreted via 
a change of time or length scale (earlier time or larger length scale),
or the region studied could be considered to be a denser than average region
in a low density (e.g., $\Omega_0=0\.1, \lambda_0=0$) universe model.}
are less useful than the change of $\xi_0$ with redshift.

\def\zmed{$z_{\mbox{\rm \small med}}$}

To the extent that Eq.~\ref{e-defneps} is a good approximation, 
direct observational estimates of $\epsilon$ at varying 
{\zmed} (median redshift) include 
$\epsilon=1\.6\pm0\.5$ (\cite{WIHS93}~1993, {\zmed}$=0\.4$), 
$\epsilon=-2\.0\pm2\.7$ (\cite{CEBC}~1994, {\zmed}$=0\.16$), 
$\epsilon\approx2\.2$ (\cite{Shep97}~1997, {\zmed}$=0\.37$) 
and
$\epsilon=2\.8$ [cf. \S 4.1(3), \cite{CFRS-VIII}~(1996), 
adopting $r_0=5\.0 h^{-1}$~Mpc and $c${\zmed}(Stromlo-APM)$=15,200$\k  
(\cite{Love92}~1992, 1995), {\zmed}(CFRS)$=0\.56$].
By hypothesising major
changes in visible galaxy populations from low to high redshifts, 
these values of $\epsilon$ could be lowered.

Simple theoretical values of $\epsilon$ include
those expected for clustering fixed in comoving coordinates 
($\epsilon=\gamma-3 \approx -1\.2$, 
so that the index in Eq.~\ref{e-defneps} is zero); 
clustering fixed in proper coordinates on small
scales [``stable clustering'', $\epsilon=0$; since the numbers of 
``clusters'' changes by $(1+z)^3$ and the number of galaxy pairs 
by $(1+z)^6$, the factor in Eq.~\ref{e-defneps} for $r$ in proper 
coordinates is $(1+z)^{-3}$]; 
and linear growth of initial fluctuations for an $\Omega_0=1$ 
universe ($\epsilon=\gamma-1 \approx 0\.8$).  

\begin{table}
\caption{ \label{t-eps} Low redshift spatial correlation function growth rates 
$\epsilon$ (see Eq.~\protect\ref{e-defneps}). Only the four lowest redshift 
points are used for $n=-2.$}
$$\begin{array}{ c cc }
\hline  \rthresh \rthstrut  & n=0 & n=-2 \\ \hline
5    & -0\.6\pm0\.1   & -0\.6\pm0\.6 \\
1000 &  -0\.4\pm0\.3  & -0\.3\pm0\.3
\\ \hline
\end{array}$$
\end{table} 

More detailed analyses include those of \cite{Peac96}~(1996) and
\cite{Matar97}~(1997), who discuss 
departures from the power law of Eq.~\ref{e-defneps}, bias factors 
and differences between the linear, quasi-linear and non-linear overdensity
regimes. In particular, the transition between the linear and non-linear
regimes can give rise to $\epsilon \approx 2\.8$ [for $\gamma=1\.8$ in 
Eq.~(29) of \cite{PD96}~(1996) and $\Omega_0=1$].

By contrast to the above estimates, the values of $\epsilon$ for our
haloes are those for which {\em haloes merge between time steps}. This means
that as haloes approach each other and merge, halo pairs are replaced by 
single haloes, so that the correlation decreases more slowly than if the
haloes' identities were kept separate. Because of this, the values of 
$\epsilon$ derived here (Table~\ref{t-eps}) lie between those 
expected for clustering fixed in comoving and 
proper coordinates respectively, and 
are lower than the more precise of the observational estimates.

Implications of the effect of merging, in particular on the 
angular correlation function, have been presented in more detail in 
\cite{RY93}~(1993).

Another property to note is that for the $n=-2$ model, 
$\xi_0$ initially {\em decreases} until $z\sim 1,$ after which it increases 
to the present.
This is because during the transition from linear perturbation amplitudes to
non-linear collapse, small length-scale perturbations on top of large
length-scale perturbations have their density boosted, and so can collapse
well before small length-scale perturbations in low density regions. 
During this
period, perturbations in the low-density regions are not 
represented by collapsed
objects.  Hence, the mean number density of collapsed objects used to normalise
the correlation function represents
the numbers of haloes in only the high-density regions, 
but includes the volume in both low
and high-density regions. The mean density is therefore biased low, 
so the correlation function amplitude is biased high. As a larger fraction of
perturbations collapse, this initially high bias in $\xi$ decreases until
normal correlation growth takes place.

	This effect is not seen for $n=0$ because large length-scale 
perturbations have relatively less power for $n=0$ than for $n=-2.$
Possible consequences of such a ``Decreasing Correlation Period'' 
(also reported by \cite{Rouk93}~1993 and \cite{BV94}~1994) on 
observations of the angular correlation function are discussed in
\cite{ORY97}~(1997). From Fig.~\ref{f-xi0evol}, 
the rate of correlation decrease 
is $d(\log_{10}\xi_0)/d[\log_{10}(1+z)] \sim 0\.3\pm 0\.15$ to 
$z\sim 0\.5-1.$

\finterp

\subsubsection{Merging Between Output Time Steps} \label{s-interp}
The merging history trees of haloes could in principle be analysed with
a higher time resolution than in the figures presented here, merely by
using many more output time steps from the N-body simulation. The only
drawback would be practical handling of computer disk space. 
A single time step for the \cite{Warr92}~(1992) simulations normally
occupies 32~Megabytes. 

Here we note an alternative to using more output times. 
Interpolation between time steps, making analytical 
assumptions for the trajectories of the haloes, can give interpolated
merger times which are statistically consistent with the more accurate 
trajectories calculated in the full N-body simulation. 
Fig.~\ref{f-interp} shows that the approximation of point particles falling
into isothermal potentials gives an approximately decreasing merger rate,
which is consistent with the overall merging rate.  
A small fraction of objects have estimated
merger times greater than the output time step interval, i.e., greater
than the time required for a merger 
according to the full N-body simulation. This is a limitation 
to (at least this) analytical interpolation, but the fraction of such 
cases is small enough that this interpolation technique may 
be a fair approximation statistically.

However, N-body analyses of few-galaxy systems (e.g., 
\cite{PrugC92}~1992; \cite{HernW89}~1989) 
suggest that merger times may be difficult to predict by any simple
analytical method, so caution is needed. Although the results should
not differ significantly from those of the authors just cited, use
of such interpolations for the calculation of N-body merger history
trees could be verified by comparison with the intermediate time steps
of the N-body systems.	Interpolation is not adopted for this paper.

\section{Galaxy Formation Application (Star Formation and Evolution)} 
\label{s-other}

The example of a galaxy formation recipe adopted here combines 
(a) gravity: halo merging history trees are derived from 
density peaks detected above a given overdensity threshold 
in $1\.1$ million particle pure gravity N-body simulations;
and
(b) other physical processes: 
one galaxy is assumed to form in each halo; 
we insert an observationally
inspired star formation rate due to merger-induced starbursts (and/or
a quiescent star formation rate); and we
combine this with the initial mass function and stellar evolution
modelled in Bruzual's (1983) evolutionary population synthesis code.

\subsection{Modelling Starbursts to Occur on 
		Merging} \label{s-bursts}  
	Star formation physics is represented by an initial mass function
(IMF) and a star formation rate (SFR). 
The separability of these two functions 
appears to be a reasonable assumption according to both 
observation and theory, e.g., the recent ISO (Infrared Satellite Observatory)
observations of both infrared-ultra-luminous and ``normal'' star-forming
galaxies are consistent with existence of a universal IMF 
(\cite{Lutz96}~1996; \cite{Vigr96}~1996; \cite{GG96}~1996).
The IMF is discussed in \S\ref{s-geps},
while this section describes the modelling 
of the SFR, based on starbursts.

      For the purpose of this first-order examination of the effect of
merger-induced starbursts on galaxy luminosity evolution, we only use very
simple models of the starbursts. Observational and 
theoretical motivations are as follows.

      An early observational model of a starburst (not necessarily 
caused by a merger), is that of Rieke \etal \ (1980).

      Rieke \etal \ (1980) model the starburst in the nucleus of M82
via evolutionary population synthesis. They find that instantaneous
(Dirac delta function) and constant star formation rate models fail to produce
the observed spectrum, but exponential decay SFR models with an IMF with
a low-mass cutoff well above a solar mass are necessary. Their best models
(say, D and F) have the e-folding time in the SFR $t_0=2\e{7} $yr and 
$t_0=1\e{8}$yr and run for $t=5\e{7} $yr and $t=1\.6\e{7} $yr respectively.
Both have IMF's with $\alpha=2$ and the mass range $3\.5 - 31 M_{\sun}.$
The mass turned into stars is $\approx 1\.5 - 2 \e{8} M_{\sun},$ this being
constrained to be less than the total mass in the nucleus, estimated as
$3\e{8} M_{\sun}$ by Rieke \etal \ This constraint is the
major reason for the
need of a high low-mass cutoff. If a normal
low-mass cutoff is used, then the mass actually present
in the nucleus is insufficient to generate the observed luminosity. 

      Rieke (1991) describes more recent observational constraints
on the models for M82, finding that the above conclusion still holds using
more modern stellar evolutionary tracks in the models.

      Scoville \& Soifer (1991) argue from IRAS far-infrared data that
``virtually all of the strong {\it global} starbursts occur in ...
starburst-infrared galaxies,'' where the latter are 
defined as ``those with $L_{IR}/M_{H_2}$ 
significantly higher than in normal galaxies,'' and that 
starburst-infrared galaxies correlate
highly with ``the occurrence of a recent [galaxy-galaxy] interaction.''
They argue that such global starbursts require the progenitor galaxies 
to be of
comparable mass in order to generate such activity.

While this result 
doesn't necessarily imply the converse, i.e., that all mergers of
similarly sized galaxies
induce major global starbursts, the converse is a fair hypothesis.
With the assumption of this converse, 
Scoville \& Soifer find that the high luminosity end
of the infrared (galaxy) luminosity function from the IRAS survey is 
consistent with $0\.2\%$ of all spiral galaxies undergoing global starbursts
at the present with lifetimes equal to the dynamical times of large galaxies,
$\sim 1-2\e{8} $yr. For the most IR-luminous
galaxies they find SFR's of $10-100 M_{\sun} $yr$^{-1}.$ They don't find a
high low-mass cutoff necessary for their models to fit the observations.

      Norman (1991), citing the models of merging gas-rich disk galaxies of
Hernquist and Barnes, (\cite{Barn90}~1990, equal mass galaxies; 
Hernquist 1989, differing mass galaxies), describes a qualitative
three-phase model to
take into account gas falling into the galaxy centre. The three phases 
essentially
correspond to proportions of the gas having fallen to the centre. Published
star formation models following these three phases separately are not
cited in the article, but would obviously be of interest.

      Norman (1991) also argues that constant SFR models of starbursts
satisfy an observed
 sparsity of post-starburst galaxies relative to starburst galaxies,
but that instantaneous SFR models predict too many post-starbursts.

      Hence, for this example application of an N-body derived
halo merger history tree, a starburst with a constant SFR
is chosen.

        For pairwise mergers, the following canonical values are chosen.
We normalise the rate of the starburst for a ``typical''
large galaxy merger product
to be an SFR of 
$\psi_0= 50 M_{\sun} $yr$^{-1}$ 
as in the models of Scoville \& Soifer (1991).
The low-mass cutoff in the IMF used here ($0\.08 M_{\sun}$, 
see Eqn~\ref{e-imf}) is consistent with Scoville \& Soifer's
value of $0\.1 M_{\sun}.$

We consider the merger product to be the remnant of two large galaxies each of
 gas mass 
$\Mgas_0= 10^{10}M_{\sun},$ 
total mass $\Mtot_0= 10^{12} M_{\sun}$ and halo 
radius $50 \,$kpc.
This gives a dynamical time $\tdyn\equiv (G\rho_0)^{-1/2} \approx 2\e{8} $yr,
where the mean density of either galaxy to its
halo radius is $\rho_0= 8\e{6} M_{\sun}$kpc$^{-3}.$
      The modelling by 
\cite{Barn90}~(1990), Hernquist (1989) or earlier non-gaseous N-body models
such as those of Quinn \& Goodman (1986) find that the merger takes place over
only a few dynamical times. 
So the choice of progenitor galaxies here gives a $\tdyn$ matching
Scoville \& Soifer's 
burst duration of $\tburst_0= 2\e{8} $yr, which is chosen
as the canonical burst duration.

      In this canonical case, $50\%$ of the total gas mass is used up in 
the burst. To sum up, we have

\begin{eqnarray}
\psi_0 &=& 50 M_{\sun} \mbox{\rm yr}^{-1} \cr 
\Mgas_0 &=& 10^{10}M_{\sun}               \cr 
 \Mtot_0 &=& 10^{12} M_{\sun}             \cr 
  \rho_0 &=& 8\e{6} M_{\sun} \kpc^{-3}    \cr 
 \tburst_0 &=& 2\e{8} \mbox{\rm yr.}
\label{e-canon}
\end{eqnarray}

The canonical values are then scaled for haloes of arbitrary mass. 
To first order
it seems reasonable that the kinetic energy 
available for generating star formation should be proportional to the mass
of the smaller halo and that the SFR should also be proportional to the
total amount of gas available. 
So, the SFR is scaled by the mass of the smaller halo times the ratio of
combined gas mass to $ 2 \Mgas_0$. 

Since we consider the 
duration of the starburst to be the order of a dynamical time, 
$\tburst$ is scaled
by $\rho^{-1/2},$ where $\rho$ is the density of the larger halo.

Given a coarse time resolution in the merging histories, each merger is 
identified by the code as a multiple merger---e.g., seven haloes merge to
one---instead of as a series of several individual pairwise mergers.
If we consider the approximation that each of the pairwise mergers involves
the largest progenitor halo, then 
a compromise can be carried out as follows.

      (a) Have a single burst with
the above normalisation.
Scale the SFR by the sum of the masses of each of the smaller
haloes (i.e., all but the largest) and by the ratio of the combined gas
mass from all progenitor haloes 
(i.e., including the largest) to $2 \Mgas_0,$ giving
\begin{equation}
   \psi(t)=\psi_0  
      {\sum_{i\neq  \imaxtiny} \Mgal_i \over \Mtot_0}
        {\sum_{{\rm all}\,i} \Mgas_i \over 2 \Mgas_0}
\label{e-psi}
\end{equation}
where $\psi(t)$ is the star formation rate (in $M_{\sun} $yr$^{-1}$), 
the progenitor
haloes are labelled by $i$, and $\imax$ is the label of the progenitor
halo of greatest mass. (For the group-finding algorithm 
used in this article, $\imax=1.$)

(b) Scale the starburst duration as above,
by $\rho^{-1/2},$ where $\rho$ is the density of the largest halo,
giving
\begin{equation}
  \tburst=\tburst_0 \left({\rho\halo\over\rho_0}\right)^{-1/2}.
\label{e-tburst}
\end{equation}
Two modifications may have to be applied in some situations.
Firstly, where the starburst at this rate uses up more gas than is actually 
available, it is truncated at the point of time when zero gas mass is left.
Secondly, if the duration of the starburst is longer than
the time interval to the next time step, it is truncated at that next
time step.

 While this modelling of multiple
merger-induced starbursts with large time steps
may make the luminosity evolution more temporally discrete than it should be,
it does conserve physical quantities in line with the 
observational and theoretical constraints discussed above, and 
should be sufficient for this demonstration of the use of N-body based
merger history trees.

\subsection{Connection with Stellar Evolutionary Population 
Synthesis}  \label{s-geps}  
	We use stellar evolutionary population synthesis to combine
star formation and stellar evolutionary physics.
      A version of Bruzual's stellar evolutionary population
synthesis code (Bruzual 1983) which is 
essentially that of 1983, but with some updating
and conversion to Unix, is the primary population synthesis 
code used, but the code of \cite{GRV87}~(1987) and 
\cite{RVG88}~(1988) (1993 version) gives similar results.  
In the interactions of the programs 
presented above with this
code, the SFR history 
and masses of galaxies and gas are determined outside of the
Bruzual routines or by amended versions of the Bruzual routines.
The return of gas from supernovae to a galaxy was turned off for test 
purposes but otherwise left on. The loss of this supernova gas from a galaxy
was not invoked, neither was the option allowing 
infall of gas to a galaxy. 

	The initial mass function (IMF) used was the default
IMF chosen by the code, after Scalo (1986) (e.g., Fig.~16 in Scalo, 1986).
Where $f(m) \propto m^{-(1+x)}$ is the number of stars born per 
unit (linear) mass in a given mass range, the IMF slopes used are
\begin{equation}
       x = \left\{ 
	\begin{array}{lll}
	-2\.60  , &  0\.05 \le M \le    0\.08 M_{\sun} 
		&\mbox{(brown dwarfs)} \\
       -2\.60  ,  & 0\.08 \le M \le    0\.18 M_{\sun} \\
        0\.01  ,  & 0\.18 \le M \le    0\.42 M_{\sun} \\
        1\.75  ,  & 0\.42 \le M \le    0\.62 M_{\sun} \\
        1\.08  ,  & 0\.62 \le M \le    1\.18 M_{\sun} \\
        2\.50  ,  & 1\.18 \le M \le    3\.5 M_{\sun} \\
        1\.63  ,  & 3\.50 \le M \le   75 M_{\sun}. 
	\end{array}
	\right.
\label{e-imf}
\end{equation}

The Bruzual code normally works 
by using simple analytical expressions for the SFR history, so
that no numerical effects 
(e.g., rounding errors) can be introduced at this stage.
 Numerical effects can of course be present 
when the galaxy spectral energy distribution is calculated, since only
finitely many points representing different stellar ages are present for each
of the finitely many mass tracks. 

      However, this has the disadvantage that one
cannot simply stop the code after a certain time step, 
save the population data,
start up the program from scratch, read in the saved population data and 
continue on as if the program had never stopped.
The population data could be stored and later read back in, 
but this would round each star's
age to the appropriate stellar evolution track age at every time step, making
cumulative errors. 

Bruzual's code was therefore modified slightly in order 
to allow use of numerical SFR 
histories. For each time step and each halo, 
an array of time points from that time step
to the following time step and the corresponding
array of integrals of the SFR are stored. 
These integrals of the SFR are the total number of 
stars created since the first star
formation in any of the progenitor peaks which end up in the present peak
being worked on. 
Because
the integral of the SFR is used,
the errors are not cumulative, and in the present
version are $\sim 0\.1\%.$ 

      The program which applies 
stellar evolutionary population synthesis to the merging histories
stores an SFR history 
for each peak as it is evolved to
the next time step, adds these together for merged peaks, and from that point
on evolutionary population synthesis applies just as it would for 
an isolated galaxy, except that the merging history tree information is
contained implicitly in the complex shape of the SFR history.

      For the present model, population synthesis is applied 
by optionally having an exponentially decreasing SFR between mergers
and optionally having starburst SFR's commencing at each merger.
Gas masses and total masses are by default
conserved, i.e., the gas and total masses of a galaxy are the
respective sums of the gas and total masses of predecessor galaxies,
except that  matter accreting directly onto the density peaks
is added as gaseous mass. 
If both exponential and burst
star formation are turned on simultaneously (probably the most realistic
model)
the SFR's are simply added together, conserving the number of stars created.

\subsection{Luminosity Functions} \label{s-galstats}

This combination of N-body generated merging
history trees with other elements of a simple galaxy formation recipe 
is sufficient to generate 
luminosity functions which are comparable to the estimated 
present-day luminosity function, and so could be extended to 
an exploration of 
galaxy formation parameter space, in which the parameters are 
free rather than motivated by observation as in the present work.

The merging history trees presented above are derived from power law
initial spectrum N-body simulations intended for comparison of 
the relative properties of haloes of different masses, so need to
be renormalised in a cosmological context. For simplicity, 
an increase in the length scale by a factor of two is the 
renormalisation adopted. 
Masses scale as the cube of the length in order to
leave the numerical operations in the N-body simulations
and the merger history tree analysis unchanged and time is not rescaled.
This brings the {\em number} density of the haloes close to that of
observed galaxies. 

\fsfr

\fmlratio

A baryon fraction of 10\% is adopted. 
Nucleosynthesis results (e.g., \cite{Walk91}~1991) and 
$H_0$ estimates (\cite{Tanv95}~1995; \cite{Kundic97}~1997) make this 
a reasonable round number for a critical density universe with a null 
cosmological constant. [Some other estimates of similar baryon fractions
include Galaxy estimates of $0\.7$ out to a Holmberg radius, or down to possibly 
$0\.07$ for the whole Galaxy (\cite{Free87}~1987); $0\.25$ for galaxy 
clusters (e.g., \cite{Sara87}~1987); and $\sim 0\.1$ over the whole range 
of cosmological structures from dwarf spheroidals to rich clusters 
(\cite{Blum88}~1988).]

\flfevsb

All baryonic matter is assumed to be potentially star-forming material.
 
An example of an SFR 
history (for the galaxy resident in the most massive final halo 
in the $n=-2,\, \rthresh =5$ exponential-plus-burst model)
is shown in Fig.~\ref{f-n0b.r5n6.eb.sfr}. Both the exponentially declining 
``quiescent'' star formation rates and the merger-induced bursts are
clearly visible. Bruzual's 
SFR parameter has the value $\mu=0\.15$ 
[$\mu$ is the proportion of gas
turned into stars in a (non-merging) galaxy within $1 \,$Gyr].

Mass-to-light ratios, $\MMlum/L\IIIaJ,$ 
(using rest frame values of $L\IIIaJ$, i.e., no K-corrections) 
for the galaxies in this model at different time steps 
are shown in Fig.~\ref{f-mln0b.eb}. These mass-to-light ratios are
somewhat high relative to optical galaxy estimates 
(e.g., $2 \ltapprox \MM\disk/{L_{B\disk}} \ltapprox 7 \msls$
derived from the inner part of the optical/HI rotation curves of disk galaxies
without bulges, \cite{Free87}~1987; 
stellar population values, \cite{LT78}~1978) but similar to those estimated 
from X-ray emission in ellipticals 
($20 \ltapprox \MMtot/L_B \ltapprox 30 \msls $ within a radius $r\sim30-40$~kpc,
\cite{Caniz87}~1987).

The corresponding luminosity functions for the galaxies in this 
model are shown in Fig.~\ref{f-lfn-2b.eb}, while luminosity functions 
for either exponentially
decaying or burst SFR's only (not both) are shown in 
Figs~\ref{f-lfn-2b.e} and \ref{f-lfn-2br5.b} (for $n=-2,\, \rthresh =5$).
A local luminosity function parametrised as a Schechter function 
\begin{equation} \label{e-schech}
dN/dL = \phi^* (L/L^*)^\alpha \exp(-L/L^*) d(L/L^*) 
\end{equation}
(\cite{Sch76}~1976) where 
$\phi^*= 1\.56\e{-2} h^3$~Mpc$^{-3}$, 
$M^* =-19\.6+5\log_{10}h$ and $\alpha =-1\.1$
(\cite{EEP88}~1988), is shown for comparison. 
[This estimate is similar to the more recent Stromlo-APM 
estimate of \cite{Love92}~(1992), while the CfA2 estimate of 
\cite{Marz94a}~(1994) differs from both in having a value of
$M^*$ about $0\.7$~mag fainter.]

The luminosity functions for the 
$n=-2,\, \rthresh =5$ exponential-plus-burst and exponential-only 
SFR's are similar to one another 
(Figs~\ref{f-lfn-2b.eb}, \ref{f-lfn-2b.e}),
as the rate of the exponential SFR alone is enough to use up most of the gas,
leaving little possibility 
for the bursts to make any difference between galaxies of differing merging
histories. 
The luminosity function 
slopes are steeper than the observational faint end slopes (for field galaxies), 
and are similar to the 
steep slopes expected from the mass functions
(Figs~\ref{f-mfn-2br5} - \ref{f-mfn-2br1000}). 
The other three combinations of $n$ and $\rthresh$ for the 
exponential-plus-burst and exponential-only models give 
similarly steep luminosity functions. 
This result is similar to that of other semi-analytical models 
(e.g., \cite{KWG93}~1993). 

The ``bump'' at the bright end of some of the later luminosity functions 
(particularly in Fig.~\ref{f-lfn-2b.e}) combined with the steep slope is
suggestive of luminosity functions in clusters, but a cluster environment
is one in which merger-induced star bursts could be expected to be the more 
important rather than the less important star formation mode, 
so it is not clear whether or not 
this is a useful characteristic of the models.
	
The faint end sudden drop 
(at $M \ltapprox -16$ in these two figures) is simply due 
to the resolution limit of the models.

  The $n=0,\, \rthresh =5$ burst-only model (Fig.~\ref{f-lfn0br5.b}) 
is quite striking in having a ``faint end'' slope almost precisely 
as shallow as the slope of the observational luminosity function.
In addition, 
the bright ends of the earlier of these luminosity functions 
have steep drops as in
the observational luminosity function, though this occurs 
for galaxies a few magnitudes too bright. 

How significant are the luminosity functions of the 
$n=0,\, \rthresh =5$ burst-only model? 
For a flat rotation curve for the Galaxy with 
density falling off as $r^{-2},$ 
the detection threshold of $\rthresh =5$ corresponds to 
a Galaxy halo radius of about $1500 \,$kpc, larger than
any claimed value. Apart from the recent 
observational suggestion 
of \cite{HonSof96}~(1996) that the halo is only 15~kpc in radius, a 
conventional estimate of $r_{halo}\sim 100-300$~kpc would still imply that the
detection threshold used here overestimates the mass of the Galaxy by
an order of magnitude [since $\rho \propto r^{-2} \Rightarrow M(<r) \propto r$].
Also, members of the Local Group would missed, since one galaxy per halo
and maximal merging are assumed.

These problems 
could conceivably be corrected by a reduction in galaxy masses, which would 
reduce the luminosities, 
and by an increase in the normalisation to account for the missing Local Group 
galaxies, whose analogues would presumably be detected (in the model) in 
isolation in other regions of space. However, while 
the former would improve the fit, 
the latter would worsen it.

	The $n=0, \rthresh =1000$ burst-only 
model does include these smaller galaxies 
(``halo substructure'') and results in a luminosity function which is 
slightly too steep (Fig.~\ref{f-lfn0br1000.b}). 
The $n=-2$ models (Figs~\ref{f-lfn-2br5.b}, \ref{f-lfn-2br1000.b}) 
also have slopes which are relatively shallow, but none as shallow 
as that for the $n=0,\, \rthresh=5$ model. 

The reason why 
the burst-only models have shallower slopes can be explained 
fairly simply: smaller mass haloes typically undergo less merging, so 
become relatively underluminous relative to 
higher mass haloes. {\em Star formation dominated 
by merger-induced bursts could therefore provide a simple way of reducing 
the faint-end slopes expected from the Press-Schechter function.}

\flfnrthr

As might be expected from the similarity
in the mass functions (Figs~\ref{f-mfn-2br5}, \ref{f-mfn-2br1000}, 
\ref{f-mfn0br1000})
for three of the burst-only models, the
luminosity functions are also similar to one another. 
However, a characteristic of the
mass function of the $n=-2,\, \rthresh=5$ model appears (marginally) to
be shared by the corresponding luminosity function: the slope seems
to become shallower with time. If this were significant, it could be of
considerable interest in reconciling faint galaxy counts 
(e.g., \cite{TSei88}~1988; \cite{Tys88}~1988) with the flatness
of the locally estimated luminosity function, though it would also need
to show evolution in the bright end of the luminosity function as
deduced from recent redshift surveys 
(\cite{CFRS-VI}~1995, \cite{Ell96LF}~1996 and \cite{Cow96}~1996). 

The evolution in slope could be attributed to 
the evolution in the Press-Schechter mass function shown in 
Fig.~\ref{f-mfn-2br5}, with the addition of a systematic reduction 
in slope induced by our burst-only SFR formalism. 

	Both the flatness of the $n=0,\, \rthresh=5$ burst-only 
luminosity functions and the decreasing slope of the
$n=-2,\, \rthresh=5$ model suggest that these parameter combinations 
should be interesting for future explorations of 
galaxy formation models using N-body derived merger history trees.

\section{Conclusion} 
	The algorithms and results discussed above show that this 
method of deriving halo merging history trees directly from N-body simulations
is feasible and easily combined with evolutionary stellar population synthesis
for exploration of a simple galaxy formation model. 
The method has been applied to \cite{Warr92}'s (1992) N-body simulations
with $n=0$ and $n=-2$
power law initial perturbation power spectra, using $\rthresh =5$ and 
$\rthresh =1000$ overdensity detection thresholds to detect dark
matter haloes.

Several subsets of the merging history
trees have been plotted, directly showing quantitative 
patterns of halo merging from fully non-linear calculations.
The merging history trees were 
combined with simple, observationally normalised proportionality assumptions
for star formation rates and stellar 
evolutionary population synthesis in order to
demonstrate a simple galaxy formation ``recipe'', in which star formation
is either exponentially decreasing, induced by mergers, or both.
In addition, interesting properties of halo formation and evolution
were noted.

The galaxy formation recipe adopted shows that 
if most star formation occurs as merger-induced bursts, then 
some flattening of the faint end 
of the galaxy luminosity function relative to 
that expected from the mass functions may be obtained. Indeed, 
for the analysis using 
(a) a white noise perturbation spectrum slope ($n=0$)
and 
(b)  a detection threshold typical of that for a perturbation
just reaching the turn-around density
($\rthresh =5$),
the resulting present-day luminosity function  
has a faint-end slope similar to $\alpha=-1\.1,$ i.e., 
that estimated for local, field, high surface brightness galaxies. 
Moreover, for the same value of
$\rthresh$ and a perturbation slope similar to that of CDM, there marginally
appears to be a steepening of the faint end slope for increasing redshift.

However, since this is only a simple test application of merging history
trees to galaxy formation modelling, and the N-body simulations are
designed for the study of relative rather than absolute halo properties,
these results should be taken as indicative of worthwhile galaxy 
formation ``recipes'' to explore further, rather than as a
definite explanation for a luminosity function with a flat faint end.

	Other interesting by-products from this method of analysing
non-linear gravitational collapse 
are properties of hierarchical halo formation.

(1) Individual merger rates
can be very different from average merger rates and the fraction
of mass coming from accretion can be quite high. 
For example, for $n=0,\, \rthresh =5$, 
the mean number of haloes which collapse 
at any time stage and end up in a halo at the final time stage 
is $7\.4,$ the standard deviation in this quantity is 
$20\.7$ (Table~\ref{t-nances}) and 
the maximum is 233.
For the same halo detection parameters, 
a mean fraction of 32\% of the mass in the final haloes
comes directly from accretion of uncollapsed matter (Table~\ref{t-rainfrac}), 
but this fraction varies widely between different haloes.
(For the other three parameter choices, 
the merger ratios are about a factor of two lower, 
while the accretion percentages are about the same.)

(2) Low mass haloes may either form at very recent cosmological epochs
or may undergo no mergers at all throughout a Hubble time. The former
may be counter-intuitive, but can be 
simply understood as being due to the existence of 
small amplitude, small length-scale perturbations at early epochs, 
expected for a zero-mean Gaussian amplitude distribution. 
Some (if not many) perturbations have small initial
amplitudes: these only become
non-linear recently. Thus, observations such as that of 
a young, low redshift, low metallicity dwarf galaxy pair (\cite{Lipo97}~1997)
can be explained naturally within a hierarchical halo formation model.

(3) If there is sufficiently more power on large than small scales, 
e.g., $n=-2,$ then the first fluctuations to collapse (forming haloes) 
are only those 
inside of large length-scale
perturbations, so that the spatial distribution of the haloes is initially
much less uniform than that of the linear (and non-linear) fluctuations
in general. In other words, the amplitude of
the spatial correlation function of the haloes 
may start from a highly biased value (relative to the linear perturbation 
theory expectation), 
and decrease in bias until a transition redshift when ``normal'' correlation 
growth commences. Observational 
consequences of such a ``Decreasing Correlation Period''
are discussed in \cite{ORY97}~(1997).

Many further developments of the use of N-body derived 
merger history trees are obvious: different assumptions for
processes (2)-(5) cited in the introduction could be adopted; a mechanism
(e.g., dissipation) could be chosen 
to change (6), so that galaxies would not
merge every time their parent haloes merged; or at the expense of using
more computer disk space, N-body simulations with higher $N$ or 
larger numbers of output 
time steps could be used. Use of more physically complex 
group-finders (e.g., \cite{GB94}~1994) should also make some differences 
to detailed properties of the merger histories.
Systematic comparisons between recent analytical predictions 
(e.g., \cite{YNG96}~1996), 
Merging-Cell-Models models (\cite{RT96}~1996) and N-body derived merger history
trees could be made. An example of a different assumption for processes
(1)-(5) would be to replace the observationally inspired
parameters adopted here by free parameters which would
be varied to see how sensitive the reduction in the 
slope of the faint end of the luminosity
function is to these parameters. 

\section{Acknowledgements}
  We would like to thank Carlos Frenk, George Efstathiou, Cedric Lacey, 
Tim de Zeeuw, Walter Jaffe, 
Simon White, Gary Mamon, Peter Thomas, Masahiro Nagashima and 
Naoteru Gouda for several interesting and 
useful comments and suggestions. Thank you also to Bob Mann
for making many constructive recommendations.
  This research has been carried out with support from the 
Australian National University, from the 
Institut d'Astrophysique de Paris, CNRS, from an ANTARES grant, 
from a PPARC Research Fellowship, from a Centre of Excellence Foreign 
Visiting Fellowship at the National Astronomical Observatory of Japan,
and has made use of Starlink computing resources. 


\end{document}